\documentclass[9pt,notitlepage,a4paper]{article}

\usepackage[margin=1in]{geometry}
\usepackage{graphicx}
\usepackage{color}
\usepackage[ linktocpage]{hyperref}
\usepackage{amssymb,amsmath}
\numberwithin{equation}{section}
\usepackage{adjustbox}
\usepackage{multirow}
\usepackage{makecell}
\usepackage{cite}
\usepackage{caption}
\usepackage{subcaption}
\usepackage[dvipsnames]{xcolor}
\usepackage{tikz}
\usepackage{dynkin-diagrams}
\usepackage{makecell}

\hypersetup{
	colorlinks=true,
	linkcolor=blue,
	citecolor=RedViolet,
	urlcolor=BrickRed,
}

\newcommand{\Tr}{\mbox{Tr}}

\begin{document}
	\begin{flushright}
		YITP-SB-2025-20
	\end{flushright}
	
	\vskip 2cm
	
	\begin{center}
		\LARGE \textbf{Generalized Schur limit, modular differential equations and quantum monodromy traces}
	\end{center}
	\vskip 0.5cm

\newcounter{footnoteA}
\renewcommand{\thefootnoteA}{\fnsymbol{footnoteA}}

\newcommand{\footnoteA}[1]{%
	\refstepcounter{footnoteA}%
	\begingroup
	\renewcommand{\thefootnote}{\thefootnoteA}%
	\footnotetext{#1}%
	\endgroup
	\textsuperscript{\thefootnoteA}%
}

\renewcommand{\thefootnoteA}{\fnsymbol{footnoteA}}

	\begin{center}
		{\large Anirudh Deb\footnoteA{anirudh.deb@stonybrook.edu}\par}
		\vspace{0.5cm}
		C. N. Yang Institute for Theoretical Physics, Stony Brook University, Stony Brook, NY 11794-3840, USA\
		\vspace{0.5em}
	\end{center}
	
		\begin{center}
				\vspace{0.5em}
		\bf{Abstract}
	\end{center}
	
		We explore some aspects of the generalized Schur limit, defined in\cite{Deb:2025ypl}. Based on several examples, we conjecture that the generalized Schur limit as a function of $\alpha$ solves a modular linear differential equation of fixed order, with coefficients depending on $\alpha$. We also observe in examples that for Argyres-Douglas theories of type $(A_1,G)$ with $G=A_n,D_n$, the generalized Schur limit for certain negative integer values of $\alpha$, coincides with the trace of higher powers of the quantum monodromy operator. This hints at a more general correspondence between the wall-crossing invariant traces on the Coulomb branch and the generalized Schur limit, which is related to the Higgs branch. 
	
	\newpage 
	
	\tableofcontents
	\section{Introduction and summary}
	Four dimensional superconformal field theories (SCFTs) with eight supercharges have been a theoretical laboratory for studying the interplay between physics and mathematics. The superconformal index \cite{Romelsberger:2005eg,Kinney:2005ej,Dolan:2008qi,Gadde:2011uv}, which encodes information about the spectrum of the protected operators  has been a major source of results relating to mathematical physics. One particular limit of the index is called the \textit{Schur index}, which counts certain $1/4$-BPS operators, obeying certain shortening conditions on their quantum numbers  \cite{Gadde:2011ik, Gadde:2011uv, Rastelli:2014jja}. In particular, to a 4d SCFT one can associate a 2d vertex operator algebra (VOA) and the Schur index equals the vaccum character of the corresponding VOA \cite{Beem:2013sza}. This limit has several interesting modular properties. The Schur index is conjectured to be a \textit{vector valued modular form} and solves a modular linear differential equation (MLDE) \cite{Arakawa:2016hkg, Beem:2017ooy}. 	Here we will examine properties of a limit closely related to the Schur index.
	
	Recently in \cite{Deb:2025ypl}, a double scaling limit of the superconformal index, called there as \textit{generalized Schur limit} and denoted by $\hat{\mathcal{Z}}(q,\alpha)$, was defined. This limit is parametrized by a parameter $\alpha\geq 0$ and for a subset of theories, at special values of this parameter, the limit reduces to the Schur index of theories related to the initial theory by renormalization group flows. In certain cases, this limit provides a very simple contour integral expression for  the Schur index indices of non-Lagrangian  SCFTs. \footnote{Using this limit \cite{Pan:2025vyu} computed a closed form expression for partially flavored Schur index of rank-one $E_6$ and $E_7$ Minahan-Nemeschansky theories.}.  This limit is well defined for any $\alpha\geq 0$, but it may or may not give rise to a integer coefficient $q$-series.	In this paper we explore the range $\alpha<0$. To analyze this range, we evaluate the index as a $q$-series with $\alpha$ dependent coefficients and use the expression to analytically continue to negative $\alpha$. In this range, for certain special values of $\alpha$, we find that the index equals the vacuum character of vertex operator algebras, which show up in the trace of higher powers of wall-crossing invariant \textit{quantum monodromy operator}\footnote{\cite{Cordova:2015nma} used the operator $\mathcal{O}(q)$ called the \textit{Kontsevich-Soibelman} operator \cite{Kontsevich:2008fj}, which is related to the quantum monodromy as $\mathcal{O}(q)=M(q)^{-1}$.}  $M(q)$ \cite{Kim:2024dxu}. We test this idea in the case of  Argyres-Douglas theories of type $(A_1,G)$, where $G=A_n, D_n$ . In the examples, we find that the generalized Schur partition function equals the trace of a higher power $M(q)$ \footnote{A similar limit was considered in \cite{Cecotti:2015lab} and its relation to the trace of $M(q)$ was discussed.}. To be more explicit
	\begin{equation}
		\label{eq:gsiandtm}
		\hat{\mathcal{Z}}(q,\alpha)=(q;q)^{2\mathfrak{r}}\Tr\, M(q)^{-\alpha},~\alpha\in \mathbb{Z}_{\leq 1},~\alpha>\alpha^*~,
	\end{equation}
	where $\mathfrak{r}$ denotes the \textit{rank} of the theory\footnote{rank is defined as the complex dimension of the Coulomb branch. For Lagrangian theories, this coincides with the rank of the gauge group.} and $\mathbb{Z}_{\leq 1}$ denotes integers less than $1$. From the low rank examples, it appears that	$\alpha^*$ is the least negative value of $\alpha$ for which the trace of $M(q)^{-\alpha}$ does not give rise to a convergent $q$-series. At this value of $\alpha$, $\hat{\mathcal{Z}}$ either diverges or gives a $q$-series with that does not seem to be the vacuum character of any known CFT.
	
	The two sides of equation \eqref{eq:gsiandtm} seem to connect two different quantities. The moduli space of vacua of 4d $\mathcal{N}=2$ SCFTs has two distinct branches: the Higgs branch and the Coulomb branch. The left hand side is a quantity related to the Higgs branch. This is because the Schur operators contain Higgs branch operators as a subset and do not contain Coulomb branch operators. The right hand side is a wall crossing invariant defined using the data of BPS states on the Coulomb branch of the theory. \cite{Cordova:2015nma} observed that in a large class of theories, the Schur index is to equal $(q;q)^{2\mathfrak{r}}\Tr M(q)^{-1}$, thus hinting at a relation between the two seemingly distinct branches (see also \cite{Cordova:2016uwk, Cordova:2017ohl, Cordova:2017mhb,Cecotti:2010fi,Kaidi:2022sng,Buican:2024jjl,Fredrickson:2017yka,Pan:2024hcz,Deb:2025cqr,Li:2025nhc,Buican:2015hsa} ).  Therefore the equality \eqref{eq:gsiandtm} seems to hint at  a generalization of the correspondence of \cite{Cordova:2015nma} .
	As a naive check of equation \eqref{eq:gsiandtm}, note that the conjectured central charge of the 4d SCFT obtained in the generalized Schur limit  of  \cite{Deb:2025ypl} matches that of the central charge conjectured in \cite{Kim:2024dxu} with the identification $\alpha=-N$
	\begin{equation}
		\label{eq:ccharge}
		c_{\text{2d}}^{(N)}=-c_{\text{2d}} N-2(N+1)\mathfrak{r}~.
	\end{equation}
	In order to check equation \eqref{eq:gsiandtm}, we need to compute the  generalized Schur limit for negative values of $\alpha$. We will restrict ourselves to $(A_1,G)$ type theories, with $G=A_n,D_n$ , whose index is available as a contour integral obtained from deforming a  $\mathcal{N}=1$ Lagrangian theory \cite{Maruyoshi:2016tqk, Maruyoshi:2016aim, Agarwal:2016pjo} \footnote{The indices of such theories were also obtained using other techniques. See for example \cite{Buican:2015ina,Buican:2017uka}. }. The generalized Schur limit for these Argyres-Douglas theories turns out to be equal to the generalized Schur limit of some $\mathcal{N}=2$ SQCD by a rescaling of $\alpha$ by some number $\xi$ \textit{i.e.}
	\begin{equation}
		\hat{\mathcal{Z}}_{(A_1,G)}(q,\alpha_{(A_1,G)})=\hat{\mathcal{Z}}_{\text{SQCD}}(q,\alpha_{SQCD}\, \xi)~,
	\end{equation}
	where the subscripts in $\alpha$ denote the two theories. For example, for the Argyres-Douglas theories $(A_1,A_2)$, $(A_1,A_3)$ and $(A_1,D_4)$ belonging to the rank-one Deligne-Cvitanovi\'c (DC) series, the corresponding SQCD is $SU(2)$ with four flavors and $\xi=h^\vee/6$. Here $h^\vee$ denotes the dual-Coxeter number of the flavor symmetry of the Argyres-Douglas theory.
	
	This provides us with a contour integral expression for the limit for $(A_1,G)$ Argyres-Douglas theories, but computing it for negative $\alpha$ is not straightforward. To continue to negative values of $\alpha$, the most straightforward approach is to keep $\alpha$ as a parameter and expand the integrand as a $q$-series, and finally integrate each coefficient of the $q$-series. This gives a $q$-series with $\alpha$ dependent coefficients. This expression can now be evaluated for negative values of $\alpha$.
	
	Although, the process is simple, it becomes difficult for higher rank theories. One can resort to an alternative method of performing numerical fits to the coefficients of the $q$-series with $\alpha$ variable. The integral expressions for SQCDs are easy to compute for a set positive integer $\alpha$.  We perform a numerical fit for the $q$-series coefficients over this set of points. This approach is also good but it becomes difficult as one wants to fit coefficients of higher powers of $q$.
	
	This brings us to another approach of using the MLDE. As mentioned above, the Schur index solves an MLDE. A finite number of numerical coefficients are sufficient to define an MLDE and therefore the Schur index can be computed upto arbitrary high orders only with this finite amount of data. In \cite{Deb:2025ypl}, it was noted that for all values of $\alpha$ where the generalized Schur limit could be computed explicitly as $q$-series expansion, the $q$-series thus obtained solved a modular linear differential equation (MLDE) of the same order. We find this to be true in all the examples we have considered. Here, we will show that this is true for the case of $SU(2)$ with four flavors, in a $q$-series expansion upto some order. For higher rank cases we check this statement for positive $\alpha$ and find it to be true in whichever value this check could be performed. Based on this, we conjecture the following
    \vspace{5pt}
    
    \textit{The generalized Schur limit for a 4d $\mathcal{N}=2$ SCFT is the solution of a fixed order MLDE, whose coefficients depend on the parameter $\alpha$.}
    \vspace{5pt}
    
    Assuming this conjecture, the generalized Schur limit of an integral reduces to finding a modular linear differential operator (MLDO) that annihilates the Schur index for all values of $\alpha$. To determine this MLDO, we will first plot the coefficients of the respective MLDO as a function of $\alpha$ and perform a fit to obtain the MLDO as a function of $\alpha$. Solving this MLDE gives the generalized Schur index as a function of $\alpha$. We use the expression thus obtained to explore the range $\alpha<0$. In most general case, the cofficients of the $q$-series and MLDE can be arbitrary functions of $\alpha$, but in all the examples we consider here, we find that the coefficients of $q$-series are rational functions of $\alpha$ and the coefficients of MLDE are polynomials in $\alpha$. As the rank of the theory increases, the order of the MLDE is also expected to increase \cite{Deb:2025cqr}. Therefore, for checks upto very low orders in $q$, the method of directly fitting the $q$-series coefficients is sometimes better and we use this technique to perform checks by refining the index by the flavor fugacities for higher rank theories.
    
    Even though we have verified the above claims in a limited number of examples, the results seem to indicate a more general structure. A concrete mathematical problem is to show rigorously that at least in the case of Lagrangian theories, where the generalized Schur limit can be taken, the corresponding MLDE is of fixed order with coefficients depending on $\alpha$. Currently, this limit can be conviently evaluated for theories which have a Lagragian description or atleast a contour integral for the index using a Lagrangian that flows to them in the infrared. It would be interesting to show this for other non Lagrangian theories. In these computations we encoutered integer coefficient $q$-series, which were not able to identify with known VOAs. It would be interesting to determine whether these correspond to a genuine VOA. 	With respect to MLDE computations, we mainly discuss the unflavored case but it would be interesting to check the properties of flavored MLDEs for the generalized Schur limit with all the flavors fugacities  \cite{Zheng:2022zkm, Pan:2023jjw}.
    
    We only do the examples related to $(A_1,G)$ type Argyes-Douglas theories, but the numerical fit technique for determining MLDEs or the coefficient of $q$-series is quite general and can be easily applied to various examples. Using fixed order MLDE to  bootstrap the  characters of 2d RCFTs is an old subject \cite{Mathur:1988rx,Mukhi:2019xjy, Mukhi:2019cpu,Mukhi:2020gnj} and similar in spirit to computations in this paper. We will discuss the rank-one DC in detail in Section \ref{sec:su2nf4}, but a natural question to ask is what happens for higher rank DC series. Interestingly, the rank-two series solves a fourth-order MLDE parametrized by a single parameter. One also obtains integer $q$-series for certain rational values of this parameter, as well as the rank-two analouges of $\mathfrak{a}_{\frac{1}{2}}$ and $\mathfrak{e}_{7\frac{1}{2}}$, resembling the structure in the rank one case. But this does not coincide with the generalized Schur limit of the $USp(4)$ Lagrangian theory in that series. This seems to indicate that there might be some interesting integral formula for the higher rank DC series, but it may or may not be some limit of the full superconformal index. It would be interesting to explore this in future.
    
  \textbf{Organization of the paper}: We review the generalized Schur limit, MLDEs and quantum monodromy trace in Section \ref{sec:background}. In Section \ref{sec:su2nf4} we discuss the $SU(2)$ with four flavors in detail. We discuss higher rank cases in Section \ref{sec:higherrank}.  Appendix \ref{app:notations} collects the some notations and coefficients of MLDEs used in this paper.
  
  \textbf{Note:} While we were in the final stages of the project, we were informed about another group studying similar topics with some overlap. This paper is being submitted in coordination with their paper \cite{Chandra:2025qpv}. We thank its authors for their cooperation.

   \section{Background}
   \label{sec:background}
   In this section we review the generalized Schur limit, MDLEs and quantum monodromy traces.
   \subsection{Generalized Schur limit}
   The superconformal index of a 4d $\mathcal{N}=2$ SCFTs is given by \cite{Romelsberger:2005eg,Kinney:2005ej,Dolan:2008qi,Gadde:2011uv}
   \begin{equation}
   	\mathcal{I}(p,q,t)=\Tr_{{\mathbb S}^3}\,(-1)^F \left(\frac{p q}t\right)^{-r}p^{j_2-j_1}\,q^{j_1+j_2}\,t^{R}\,\prod_{i=1}^{\text{rank\,} G_F}u_i^{F_i}~,
   \end{equation}
   where $p,q$ and $t$ are superconformal fugacities and $u_i$ are the flavour fugacities. $R$ is the Cartan generator
   of the $SU(2)_R$ R-symmetry, $r$ is the $U(1)_r$ R-charge, and
   $j_i$ are the Cartan generators of the $SU(2)\times SU(2)$ isometry of $\mathbb{S}^3$. The limit $t=q$ is called the Schur index and in this limit, the index is independent of $p$ \cite{Gadde:2011ik, Gadde:2011uv, Rastelli:2014jja}
   \begin{equation}
   {\mathcal{I}}_\text{Schur}(q)=\Tr \,(-1)^F\, q^{R+j_1+j_2}\,\prod_{i=1}^{\text{rank\,} G_F}u_i^{F_i}\,.
   \end{equation}
   The Schur index counts Schur operators which satisfy the following shortening conditions 
   \begin{equation}
   		E-(j_1+j_2)-2R=0~,\;\;\;
   		\; r+j_1-j_2=0~.
   \end{equation}
   To every 4d $\mathcal{N}=2$ SCFT, one can associate a vertex operator algebra (VOA) \cite{Beem:2013sza}. As a vector space, the corresponding VOA captures the Schur operators and the vacuum character of the VOA equals the Schur index of the SCFT. The Higgs branch operators of the theory are a subset of the Schur operators. In fact, the \textit{Higgs branch reconstruction conjecture} asserts that the Higgs branch can be recovered from the VOA by the associated variety construction \cite{Arakawa:2016hkg, Beem:2017ooy}. Assuming the Higgs branch reconstruction conjecture, it can be shown that the VOA vacuum character must obey a monic modular linear differential equation (MLDE). Therefore Schur indices are expected to be solutions to MDLEs.
  
 Now, let us recap the generalized Schur limit. This limit is obtained by taking the limit $\epsilon\to 0$ in the following parametrization of the fugacities
  \begin{equation}
  	p=1-\epsilon,~t\to \left(q p\right)^{1+\frac{\alpha}{\log q}\epsilon}~.
  \end{equation}
  For Lagrangian theories, the limit after stripping off the pole singularity and normalizing appropriately takes the following form \footnote{We are using the following notations
  	\begin{equation}
  		(a;q)_\infty=\prod_{i=1}^{\infty} (1-a q^i),~~	(a;q)_n=\prod_{i=1}^{n-1} (1-a q^i)
  \end{equation}}
   \begin{equation}
   	\label{eq:integralexpression}
   	\hat{\mathcal{Z}}(q,\alpha)=\frac{(q;q)_\infty^{2(1-\alpha)\mathfrak{r}}\oint [d\mathbf{a}]\left(\Delta(\mathbf{a})\mathcal{I}_{\text{vec}}(\mathbf{a},q)\mathcal{I}_{\text{hyp}}(\mathbf{a},q)\right)^\alpha}{\oint [d\mathbf{a}]\Delta(\mathbf{a})^\alpha}~.
   \end{equation}
   $\mathcal{I}_{\text{hyp}}$ and $\mathcal{I}_{\text{vec}}$ denote the indices for the vector and hypermultiplet, which can be written as the \textit{plethystic exponential}\footnote{~$
   		\text{PE}\left[f(x_1,x_2,\dots,x_n)\right]:=e^{\sum_{k=1}^\infty \frac{1}{k}f(x_1^k,x_2^k,\dots,x_n^k)}$} of single particle indices as follows
   \begin{align}
   	\mathcal{I}_{\text{hyp}}(\mathbf{a},q)&=\text{PE}\left[\frac{q^{\frac{1}{2}}}{1-q}\chi^{\mathfrak{g}}_{\mathcal{R}}(\mathbf{a})\right]~,\\
   	\mathcal{I}_{\text{vec}}(\mathbf{a},q)&=\text{PE}\left[\frac{-2q}{1-q}\chi^{\mathfrak{g}}_{\text{adj}}(\mathbf{a})\right]~.
   \end{align}
   $\chi^{\mathfrak{g}}_{\mathcal{R}}(\mathbf{a})$ denotes the character of representation $\mathcal{R}$ of the half-hypermultiplets in the gauge algebra $\mathfrak{g}$.  Note that in this limit, the Haar measure $\Delta(\mathbf{a})$ is raised to the power $\alpha$. It has been observed in a variety of examples that, for certain positive values of $\alpha$, this limit equals the Schur index of theories related to this theory by RG flows. It was also conjectured in \cite{Deb:2025ypl} that a necessary condition for the partition function of two theories $\mathcal{T}_1$ and $\mathcal{T}_2$ to be related to each other as follows
   \begin{equation}
   		\hat{\mathcal{Z}}_{\mathcal{T}_2}(q,1)=\hat{\mathcal{Z}}_{\mathcal{T}_1}(q,\alpha)~,
   	\end{equation}
   	is that their 4d central charges $c_{\text{4d}}^{(1)}$, $c_{\text{4d}}^{(2)}$ and the Coulomb branch scaling dimensions $\Delta_{i=1,2,\dots,\mathfrak{r}}^{(1)}$, $\Delta_{i=1,2,\dots,\mathfrak{r}}^{(2)}$ are related as follows 
   	\begin{equation}
   		\label{eq:cdeltaconj}
   		c_{\text{4d}}^{(2)}=c_{\text{4d}}^{(1)}\alpha+\frac{1}{6}(1-\alpha)\mathfrak{r}\,,\;\; \;
   		\Delta^{(2)}_i=(\Delta^{(1)}_i-1)\alpha+1~.
   	\end{equation}
   The VOA/SCFT correspondence states that the 4d and 2d central charges are related as follows
   \begin{equation}
   	c_{2d}=-12c_{4d}~.
   \end{equation}
   Rewriting the 4d central charge in equation \eqref{eq:cdeltaconj}, in terms of 2d central charge one obtains equation \eqref{eq:ccharge} with the identification $N=-\alpha$.
   
   Starting from the integral in equation \eqref{eq:integralexpression} for an SCFT, we have observed that the order of the modular linear differential operator (MLDO) that annihilates $\hat{\mathcal{Z}}(q,\alpha)$ is independent of $\alpha$. Only the coefficients for the MLDO depend on $\alpha$ and are a polynomial in $\alpha$. We have verified this for a number of examples for integer values of $\alpha$. In theories with higher rank, performing such integrals is difficult and we assume this conjecture regarding MLDEs to be true and use it determine this limit.  
  
   As mentioned in the introduction, the procedure is straightforward. For a given Lagrangian theory, evaluate \eqref{eq:integralexpression} for several positive integer values of $\alpha$ as a series expansion in $q$. We find the MLDO that annilates the index for each of these $\alpha$. For a given value of $\alpha$, the MLDO is found by writing the most general $n$-th order MLDO with unfixed coefficients and as a $q$-series expansion the equation is solved for the unfixed coeffcients.  For every coefficient of the MLDO, we perform a polynomial fit for various values of $\alpha$. This provides the MLDO in terms of $\alpha$, which can be solved to obtain a $q$-series expansion of $\hat{\mathcal{Z}}(q,\alpha)$.
   
   Let us show this in the case of $SU(2)$ $N_f=4$ SQCD. In this case the MLDE can be parametrized with a single coefficient $\mu(\alpha)$
   \begin{equation}
   	D_q^{(2)}+\mu(\alpha) \mathbb{E}_4(q)~.
   \end{equation}
   Determining $\mu$ at several postive integer values of $\alpha$ and performing a polynomial fit, one  obtains
   \begin{equation}
   	\mu(\alpha)=-5(6\alpha+1)(6\alpha-1)~.
   \end{equation}
   Solution to this equation with the leading power $q^{-\frac{c_{2d}}{24}}$, gives
   	\begin{equation}
   	q^{-\frac{(-12 \alpha -2)}{24}}\left(1+\frac{2 (-1+5 \alpha ) (1+6 \alpha ) q}{1+\alpha }+\frac{\left(-2-3 \alpha +29 \alpha ^2+150 \alpha ^3+1800 \alpha
   			^4\right) q^2}{(1+\alpha ) (2+\alpha )}+O(q^3)\right)~.
   \end{equation}
   This way, we obtain $\hat{\mathcal{Z}}(q,\alpha)$ as a function of $\alpha$ and therefore can be evaluated for negative values of $\alpha$ as well. In Section \ref{sec:su2nf4}, we obtain the same result by performing the $SU(2)$ integral for general $\alpha$. This method is quite general and we use it to obtain $\hat{\mathcal{Z}}(q,\alpha)$ for higher rank theories in Section \ref{sec:higherrank}.
   
     As above and  in the examples to follow, the coefficients of the $q$-series are rational functions in $\alpha$ and the coefficients of MLDOs are polynomials in $\alpha$. Since fitting a polynomial is much simpler than fitting a rational function, we will fit coefficients of MLDOs. Another advantage of this approach is that the number of coefficients describing the MLDE are finite and therefore contain the information of the entire $q$-series solution. But if one is interested in only low order coefficients, fitting the respective coefficients maybe more efficient in those cases, especially if the rank of the theory is high. We will also compute the flavored limit of the index for first few orders and in those cases we will fit the coefficients of the $q$-series.

   \subsection{Trace of quantum monodromy}
   The moduli space of vacua of 4d $\mathcal{N}=2$ SCFTs have two distinct branches, the Higgs branch and the Coulomb branch. These have two seemigly different mathemtical invariants associated to them. The Coulomb branch is described the special K\"ahler geometry \cite{Seiberg:1994rs, Seiberg:1994aj} and
   as discussed above, the Higgs branch equals the associated variety of the VOA. Apriori, the Higgs and Coulomb branch seem unrelated.  In \cite{Cordova:2015nma}, a non trivial relation between the Higgs and Coulomb branch was observed where the trace of the inverse of quantum monodromy $M(q)$, was shown to be equal to the Schur index of the theory 
   \begin{equation}
   	\mathcal{I}_{\text{Schur}}(q)=(q;q)_\infty^{2\mathfrak{r}}\Tr M(q)^{-1}~.
   \end{equation}  
   The quantum monodromy $M(q)$ is defined using Coulomb branch data as follows
   \begin{equation}
   	M(q)=\prod_{\gamma\in \Gamma}^{\curvearrowleft}\Psi(q, X_\gamma)~,
   \end{equation}
   where the product is over all BPS states with $\gamma$ in the charge lattice $\Gamma$, in the phase order of the associated central charge. The function $\Psi(q, X_\gamma)$ is known as the quantum dilogarithm and is defined as
   \begin{equation}
   	\Psi(q, X_\gamma)=\prod_{i\geq 0}(1+q^{\frac{i}{2}+1}X_\gamma)=\sum_{k\geq 0}\frac{q^{k^2/2}}{(q;q)_k}X_\gamma^k
   \end{equation}
   These operators satisfy a quantum torus algebra given as 
   \begin{equation}
   	X_{\gamma_1}X_{\gamma_2}=q^{\frac{\langle\gamma_1,\gamma_2\rangle}{2}}X_{\gamma_1+\gamma_2}=q^{\langle\gamma_1,\gamma_2\rangle}X_{\gamma_2}X_{\gamma_1}~.
   \end{equation}
$\langle\cdot,\cdot\rangle$ denotes the mutual Dirac pairing  and if a BPS quiver is provided,  $\langle\gamma_i,\gamma_j\rangle$ equals the number of arrows directed from node $i$ to node $j$.
The $\Tr$ is defined as
\begin{equation}
	\Tr X_\gamma=\begin{cases}
		1,~\gamma=0\\
		0,~\gamma\neq 1
	\end{cases}
\end{equation}
If $\gamma_{f_i}$ is a flavor charge, then $\langle \gamma_{f_i},\gamma\rangle=0$ for all $\gamma\in\Gamma$. The associated element $X_{\gamma_{f_i}}$ is central and commutes with $X_\gamma$ for all $\gamma\in \Gamma$. Therefore, for a general element $\gamma$
\begin{equation}
	\Tr X_\gamma=\begin{cases}
		\prod_{i}\Tr X_{\gamma_{f_i}}^{f_i(\gamma)}&,~\langle\gamma,\gamma'\rangle=0~\forall~\gamma'\in\Gamma\\
		0&,~\text{otherwise}~,
	\end{cases}
\end{equation}
where $f_i(\gamma)$ denotes the flavor charges of $\gamma$. The trace can be put in a basis, where the trace is related to a theory dependent function of flavor fugacities $z_i$
\begin{equation}
	\Tr X_{\gamma_{f_i}}=h_i(z_1, z_2, \dots, z_{n_f})
\end{equation}
 We will focus on theories of type $(A_1,G)$, $G=A_n, D_n$. For $(A_1,G)$ theory, the quiver corresponds to a Dynkin diagram for Lie algebra of type $G$. Their BPS quivers are shown in Figure \ref{fig:bpsquivad}. If all the edges for a node point towards it, it is called a sink node and if they point away from it, it is called a source node. The sink nodes are shown with filled black circles. The quantum monodromy for these quivers takes the form
\begin{equation}
	M(q)=\prod_{I\in\text{sink}}\Psi\left(q,X_{\gamma_I}\right)\prod_{J\in\text{source}}\Psi\left(q,X_{\gamma_J}\right) \prod_{I'\in\text{sink}}\Psi\left(q,X_{-\gamma_{I'}}\right)\prod_{J'\in\text{source}}\Psi\left(q,X_{-\gamma_{J'}}\right)~.
\end{equation}
The flavor symmetries $G_F$ and flavor lattice vectors of $(A_1,G)$ theories are collected in Table \ref{tab:flavconv}. We refer  the reader to \cite{Kim:2024dxu, Cordova:2015nma, Cecotti:2015lab} for more details on computing these traces.

\begin{figure}
\centering
\begin{minipage}{0.45\textwidth}
	\centering
\begin{tikzpicture}[node/.style={circle,draw,inner sep=3pt}, scale=1]
\node[node] (g1) at (1,0)   {};
\node[node,fill=black] (g2) at (2,0) {};
\node[node] (g3) at (3,0)   {};

\foreach \i in {1,...,3}{
	\node (lab\i) at (\i*1,0.3) {$\gamma_\i$};
}

\draw[->, >=Stealth] (g1) -- (g2);
\draw[<-, >=Stealth] (g2) -- (g3);

\node (g4) at (4,0) {\dots};

\node[node] (g5) at (5,0)   {};
\node[node,fill=black] (g6) at (6,0) {};

\node (lab5) at (5,0.3)   {$\gamma_{2n-1}$};
\node (lab6) at (6,0.3) {$\gamma_{2n}$};

\draw[->, >=Stealth] (g3) -- (g4);
\draw[<-, >=Stealth] (g4) -- (g5);
\draw[->, >=Stealth] (g5) -- (g6);
\end{tikzpicture}
\caption*{$(A_1,A_{2n})$}
\end{minipage}
\hspace{0.4cm}
\begin{minipage}{0.45\textwidth}
	\centering
\begin{tikzpicture}[node/.style={circle,draw,inner sep=3pt}, scale=1]
\node[node] (g1) at (1,0)   {};
\node[node,fill=black] (g2) at (2,0) {};
\node[node] (g3) at (3,0)   {};

\foreach \i in {1,...,3}{
	\node (lab\i) at (\i*1,0.3) { $\gamma_\i$};
}

\draw[->, >=Stealth] (g1) -- (g2);
\draw[<-, >=Stealth] (g2) -- (g3);

\node (g4) at (4,0) {\dots};

\node[node,fill=black] (g5) at (5,0)   {};
\node[node] (g6) at (6,0) {};

\node (lab5) at (5,0.3)   {$\gamma_{2n}$};
\node (lab6) at (6,0.3) {$\gamma_{2n+1}$};

\draw[->, >=Stealth] (g3) -- (g4);
\draw[->, >=Stealth] (g4) -- (g5);
\draw[<-, >=Stealth] (g5) -- (g6);
\end{tikzpicture}
\caption*{$(A_1,A_{2n+1})$}
\end{minipage}
\vspace{0.5cm}

\begin{minipage}{0.45\textwidth}
	\centering
\begin{tikzpicture}[node/.style={circle,draw,inner sep=3pt}, scale=1.2]

\node[node] (g1) at (1,0)   {};
\node[node,fill=black] (g2) at (2,0) {};
\node[node] (g3) at (3,0)   {};

\foreach \i in {1,...,3}{
	\node (lab\i) at (\i*1,0.3) { $\gamma_\i$};
}

\node (g4) at (4,0) {\dots};

\node[node] (g5) at (5,0) {};

\node[node,fill=black] (g6) at (5.5,0.7) {};
\node[node,fill=black] (g7) at (5.5,-0.7) {};

\node (lab6) at (5.0,0.7) {$\gamma_{2n+1}$};
\node (lab7) at (5.0,-0.7) {$\gamma_{2n}$};

\draw[->, >=Stealth] (g1) -- (g2);
\draw[<-, >=Stealth] (g2) -- (g3);

\draw[->, >=Stealth] (g3) -- (g4);
\draw[<-, >=Stealth] (g4) -- (g5);
\draw[->, >=Stealth] (g5) -- (g6);
\draw[->, >=Stealth] (g5) -- (g7);	
\end{tikzpicture}
\caption*{$(A_1,D_{2n+1})$}
\end{minipage}
\hspace{0.4cm}
\begin{minipage}{0.45\textwidth}
	\centering
\begin{tikzpicture}[node/.style={circle,draw,inner sep=3pt}, scale=1.2]

\node[node] (g1) at (1,0)   {};
\node[node,fill=black] (g2) at (2,0) {};
\node[node] (g3) at (3,0)   {};

\foreach \i in {1,...,3}{
	\node (lab\i) at (\i*1,0.3) { $\gamma_\i$};
}

\node (g4) at (4,0) {\dots};

\node[node,fill=black] (g5) at (5,0) {};

\node[node] (g6) at (5.5,0.7) {};
\node[node] (g7) at (5.5,-0.7) {};

\node (lab6) at (5.0,0.7) {$\gamma_{2n+1}$};
\node (lab7) at (5.0,-0.7) {$\gamma_{2n}$};

\draw[->, >=Stealth] (g1) -- (g2);
\draw[<-, >=Stealth] (g2) -- (g3);

\draw[->, >=Stealth] (g3) -- (g4);
\draw[->, >=Stealth] (g4) -- (g5);
\draw[<-, >=Stealth] (g5) -- (g6);
\draw[<-, >=Stealth] (g5) -- (g7);	
\end{tikzpicture}
\caption*{$(A_1,D_{2n+2})$}
\end{minipage}
\caption{BPS quivers corresponding to Argyres-Douglas theories}
\label{fig:bpsquivad}
\end{figure}

\begin{table}
	\centering
	\begin{tabular}{|c|c|c|c|}
		\hline
		Theory & Flavor symmetry & Flavor vectors & Trace convention\\
		\hline
		\hline
		$(A_1,A_{2N})$ &$\varnothing$ & $-$ & $-$\\
		\hline
		$(A_1,A_{2N+1})$ & \makecell{
			$SU(2)$, $N=1$ \\$U(1)$, $N>1$} & $\gamma_f=\sum_{i=0}^N (-1)^{i}\gamma_{2i+1}$ & $\Tr X_{\gamma_f}=\begin{cases} a^2 ,~N=1 \\ (-1)^{N+1}a, ~N>1 \end{cases}$\\
		\hline
		$(A_1,D_{2N+1})$ &$SU(2)$ & $\gamma_f=\gamma_{2N+1}-\gamma_{2N}$ & $\Tr X_{\gamma_f}=a^2$\\
		\hline
			\rule{0pt}{28pt} $(A_1,D_{2N+2})$ &\makecell{
			$SU(3)$, $N=1$ \\$SU(2)\times U(1)$, $N>1$} & \makecell{$\gamma_{f_1}=\gamma_{2N+2}-\gamma_{2N+1}$\\$\gamma_{f_2}=\sum_{i=0}^N(-1)^i \gamma_{2i+1}$} & \makecell{$\Tr X_{\gamma_{f_1}}=b^2$\\ $\Tr X_{\gamma_{f_2}}=\begin{cases} a b,~N~\text{odd} \\ -\frac{a}{b}, ~N~\text{even}\end{cases}$}\\
		\hline
	\end{tabular}
	\caption{Flavor symmetries of Argyres-Douglas theories and the conventions used for computing traces with flavor fugacities.}
	\label{tab:flavconv}
\end{table}

	\section{$SU(2)$ $N_f=4$ series}
	\label{sec:su2nf4}
	In this section we discuss the case of $SU(2)$ with four flavors. It was observed in \cite{Deb:2025ypl}, that for certain rational values of $\alpha$, the generalized Schur partition function coincides with the Schur indices of theories in rank-one Deligne-Cvitanovi\'c (DC)
	 series. Theories belonging to this series are labelled by their flavor symmetry $\mathfrak{g}$ belonging to the Deligne-Cvitanovi\'c series of simple Lie algebras
	 \begin{equation}
	 	\label{eq:deligneseries}
	 	\mathfrak{a}_0\subset\mathfrak{a}_1\subset\mathfrak{a}_2\subset\mathfrak{g}_2\subset\mathfrak{d}_4\subset\mathfrak{f}_4\subset\mathfrak{e}_6\subset\mathfrak{e}_7\subset\mathfrak{e}_8~,
	 \end{equation}
	The theories are also known as $\mathfrak{a}_0\simeq (A_1,A_2)$, $\mathfrak{a}_1\simeq (A_1,A_3)$, $\mathfrak{a}_2\simeq (A_1, D_4)$ Argyes-Douglas theories \cite{Argyres:1995jj,Argyres:1995xn} and the $\mathfrak{e}_6$,$\mathfrak{e}_7$,$\mathfrak{e}_8$  are known as Minahan-Nemeschansky theories \cite{Minahan:1996fg, Minahan:1996cj}. $\mathfrak{d}_4$ is the only Lagrangian theory in this series and corresponds to $SU(2)$ theory with four flavors. $\mathfrak{f}_4$ and $\mathfrak{g}_2$ do not correspond to known 4d theories.  We will denote the generalized Schur partition function of these theories by $\hat{\mathcal{Z}}_{\mathfrak{g}}(q,\alpha)$. The generalized Schur partition function for   the $\mathfrak{d}_4$ theory takes the following form
	\begin{equation}
	\label{eq:su2alpha}
	\hat {\mathcal Z}_{\mathfrak{d}_4}(q,\alpha)=\frac{(q;q)^{2}}{{\mathtt{N}}(\alpha)}\oint\frac{dz}{2\pi i z} \left(\frac{\Delta(z)(q\, z^{\pm2};q)^2}{(q^{\frac12}z^{\pm1};q)^8}\right)^\alpha~,
	\end{equation}
	\begin{equation}
	 \Delta(z)=\frac{1}{2}\left(1-z^2\right)\left(1-\frac{1}{z^{2}}\right)\,,\quad{\mathtt{N}}(\alpha)=\oint\frac{dz}{2\pi i z}\Delta(z)^\alpha\,.\;\;\;
	\end{equation}
	
	\begin{table}[t]
		\centering
		\begin{tabular}{|c|c|c|c|c|c|}
			\hline
			Theory & $h^\vee$& $12c$  & $24a$ & $\Delta$ & $\alpha$\\
			\hline
			$\mathfrak{e}_8$ & $30$ & $62$ & $95$ & $6$  & $5$\\
			\hline
			$\mathfrak{e}_{7\frac{1}{2}}*$ & $24$ & $50$ & $77$ & $5$  & $4$\\ 
			\hline
			$\mathfrak{e}_7$ & $18$ & $38$ & $59$ & $4$ & $3$\\
			\hline
			$\mathfrak{e}_6$ & $12$ &$26$ & $41$& $3$ & $2$\\
			\hline
			$\mathfrak{f}_4*$ & $9$ &$20$ &$32$ & $5/2$ & $3/2$\\
			\hline
			$\mathfrak{d}_4$ & $6$ &$14$ &$23$ & $2$ & $1$\\
			\hline
			$\mathfrak{g}_2*$ & $4$ &$10$ &$17$ & $5/3$ & $2/3$\\
			\hline
			$\mathfrak{a}_2$ & $3$ &$8$ &$14$ & $3/2$ & $1/2$\\
			\hline
			$\mathfrak{a}_1$ & $2$ &$6$ &$11$ & $4/3$ & $1/3$\\
			\hline
			$\mathfrak{a}_{\frac{1}{2}}*$ & $3/2$ &$5$ & $19/2$& $5/4$ &  $1/4$\\
			\hline
			$\mathfrak{a}_0$ & $6/5$ &$22/5$ & $43/5$& $6/5$ &  $1/5$\\
			\hline
		\end{tabular}
		\caption{Values of $\alpha$ for the Deligne-Cvitanovi\'c rank-one series. $h^\vee$ denotes the dual Coxeter number. The $a$ and $c$ central charges have been resclaed by $12$ and $24$. $\Delta$ denotes the Coulomb branch scaling dimension.}
		\label{tab:delrankone}
	\end{table}
	The values of $\alpha$ for which this limit coincides with the Schur index of theories in DC  series is given in table \ref{tab:delrankone} \footnote{The table also contains the entries $\mathfrak{a}_{\frac{1}{2}}$ and $\mathfrak{e}_{7\frac{1}{2}}$, which do not have a known corresponding 4d theory.}.	This integral can be computed for a general $\alpha$ by noting the following
	\begin{equation}
		\frac{1}{\mathtt{N}(\alpha)}\oint\frac{dz}{2\pi i z}\Delta(z)^\alpha z^x=\frac{e^{i \pi  x} \csc (\pi  \alpha ) \Gamma (\alpha +1) \cos \left(\frac{\pi  x}{2}\right) \sin \left(\frac{1}{2} \pi  (2 \alpha -x)\right) \Gamma \left(\frac{x}{2}-\alpha \right)}{\Gamma (-\alpha ) \Gamma \left(\frac{x}{2}+\alpha +1\right)}~.
	\end{equation}
   Using the above expression, the integral can be perfomed order by order in $q$ and the first few orders take the following form
	\begin{equation}
	\begin{split}
		\hat {\mathcal Z}_{\mathfrak{d}_4}(q,\alpha)&=1+\frac{2 (-1+5 \alpha ) (1+6 \alpha ) q}{1+\alpha }+\frac{\left(-2-3 \alpha +29 \alpha ^2+150 \alpha ^3+1800 \alpha
			^4\right) q^2}{(1+\alpha ) (2+\alpha )}\\&+\frac{2 \left(6+11 \alpha -424 \alpha ^2-389 \alpha ^3+8080 \alpha ^4+6300
			\alpha ^5+18000 \alpha ^6\right) q^3}{(1+\alpha ) (2+\alpha ) (3+\alpha )}+O(q^4)~.
	\end{split}
\end{equation}
	Now, we can use the above $q$-series and evaluate it for any value of $\alpha$. As a series expansion in $q$, the above coincides with the following expression as a power series in $q$ \cite{Franc_Mason_2014}
    \begin{equation}
    \label{eq:sol1su2}
		K(q)^{\frac{1+6\alpha}{12}}\,_2F_1(\frac{1+6\alpha}{12},\frac{5+6\alpha}{12},1+\alpha,K(q))~,
	\end{equation}
	where\footnote{$E_{2k}$ denotes the normalized Eisenstein series $E_{2k}(q)=1+\frac{1}{\zeta(1-2k)}\sum_{n=1}^\infty \frac{n^{2k-1}q^n}{1-q^n}$}
	\begin{equation}
		j(q)=\frac{1728 E_4(q)^3}{E_4(q)^3-E_6(q)^2},~K(q)=\frac{1728}{j(q)}~.
	\end{equation}
	Since the expression \eqref{eq:sol1su2} solves a second order differential equation parameterized by $\alpha$, this shows that that integral \ref{eq:su2alpha} solves a second order differential equation for general $\alpha$. 	The MLDE solved by these takes the form
	\begin{equation}
		D_q^{(2)}-5(6\alpha+1)(6\alpha-1)\mathbb{E}_4(q)~.
	\end{equation}

	Table \ref{tab:su2} collects the values of $\alpha$ for which we obtain $q$-series with integer coefficients. Based on the $q$-series, we note down the VOA to which these vacuum characters correspond to. These coincide with the VOA characters found in \cite{Kim:2024dxu} using trace of higher powers of the quantum monodromy. To make the connection precise, for the theory with flavor symmetry $\mathfrak{g}$, we rescale $\alpha$  as follows
	\begin{equation}
		\alpha_{\mathfrak{g}}=\alpha \frac{6}{h^\vee}~.
	\end{equation}
	In this rescaling $\alpha_{\mathfrak{g}}=1$ corresponds to the generalized Schur partition function for the theory $\mathfrak{g}$. For the case of Argyres-Douglas theories in this series, this limit coincides with generalized Schur limit for the respective theory using $\mathcal{N}=1$ Lagrangian description of \cite{Maruyoshi:2016tqk, Maruyoshi:2016aim, Agarwal:2016pjo}, with the overall rescaling of $\alpha$. For $\mathfrak{g}=\mathfrak{a}_0,\mathfrak{a}_{\frac{1}{2}},\mathfrak{a}_2,\mathfrak{g}_2$
	\begin{equation}
		\hat{\mathcal{Z}}_{\mathfrak{g}}(q,\alpha)=\frac{(q;q)^{2(1-\alpha_{\mathfrak{g}}\frac{h^\vee}{6})}}{\mathtt{N}\left(\alpha_{\mathfrak{g}}\frac{h^\vee}{6}\right)}\oint \frac{dz}{2\pi iz}\Delta(z)^{\alpha_{\mathfrak{g}}\frac{h^\vee}{6}}\text{PE}\left[\alpha_{\mathfrak{g}}\frac{h^\vee}{6}\left(\frac{-2q}{1-q}\chi_\mathbf{3}(z)+\frac{8q^{\frac{1}{2}}}{1-q}\chi_{\mathbf{2}}(z)\right)\right]~.
	\end{equation}
	Let us denote the quantum monodromy for theory $\mathfrak{g}$ as $M_{\mathfrak{g}}(q)$. In this notation, the conjecture can be stated as 
	\begin{equation}
		\hat{\mathcal{Z}}_{\mathfrak{g}}(q,\alpha_{\mathfrak{g}})=(q;q)^2\Tr M_{\mathfrak{g}}(q)^{-\alpha_{\mathfrak{g}}},~\mathfrak{g}=\mathfrak{a}_0,\mathfrak{a}_1,\mathfrak{a}_2, ~\alpha_{\mathfrak{g}}\in\mathbb{Z}_{\leq 1},~ \alpha_{\mathfrak{g}}>-\frac{6}{h^\vee}~.
	\end{equation}
	At $\alpha_{\mathfrak{g}}=-\frac{6}{h^\vee}$, the generalized Schur limit has a pole and for the values of $\alpha_{\mathfrak{g}}$ higher than this, the traces match the limit of the index. There is no known four dimensional description of the $\mathfrak{a}_{\frac{1}{2}}$ case and therefore we are not able to match VOA characters obtained as multiples $\alpha=-\frac{1}{4}$ to trace of some $M(q)$.

	    \begin{table}
	    	\centering
		\begin{tabular}{|c|c|c|c|}
		\hline
		\multicolumn{4}{|c|}{$\mathfrak{a}_0\simeq(A_1,A_2)$ }\\
		\hline
		VOA&$\alpha$ & $c_{\text{2d}}$ & $q$-series \\
		\hline
	\rule{0pt}{12pt}	$\mathfrak{osp}(1|2)_1$ &$	-\frac{1}{5}$& $\frac{2}{5}$ &$1+q+q^2+q^3+2q^4+O\left(q^5\right)$\\
	\rule{0pt}{12pt} $(\mathfrak{g}_2)_1 $&	$ -\frac{2}{5}$&$\frac{14}{5}$ &$1+14 q+42 q^2+140 q^3+350 q^4+O\left(q^5\right)$\\
	\rule{0pt}{12pt} $(\mathfrak{f}_4)_1 $ &		$-\frac{3}{5}$&$\frac{26}{5}$ &$1+52 q+377 q^2+1976 q^3+7852 q^4+O\left(q^5\right)$\\
	\rule{0pt}{12pt} $(\mathfrak{e}_{7\frac{1}{2}})_1 $&	$-\frac{4}{5}$& $\frac{38}{5}$&$1+190 q+2831 q^2+22306 q^3+129276 q^4+O\left(q^5\right)$\\
		
		\hline
		\hline
		\multicolumn{4}{|c|}{$\mathfrak{a}_{\frac{1}{2}}$}\\
		\hline
		VOA&$\alpha$ & $c_{\text{2d}}$ & $q$-series \\
		\hline
		\rule{0pt}{12pt} $(\mathfrak{a}_1)_{1} $ & $	-\frac{1}{4}$& $1$ & $1+3 q+4 q^2+7 q^3+13 q^4+O\left(q^5\right)$\\
		\rule{0pt}{12pt}$(\mathfrak{d}_4)_1 $&	$	-\frac{2}{4}$& $4$ & $1+28 q+134 q^2+568 q^3+1809 q^4+O\left(q^5\right)$\\
		\rule{0pt}{12pt}$(\mathfrak{e}_7)_{1}$&	 $	-\frac{3}{4}$& $7$ & $1+133 q+1673 q^2+11914 q^3+63252 q^4+O\left(q^5\right)$\\
		\hline
		\hline
		\multicolumn{4}{|c|}{$\mathfrak{a}_1\simeq(A_1,A_3)$} \\
		\hline
		VOA &$\alpha$ & $c_{\text{2d}}$ & $q$-series \\
		\hline
		\rule{0pt}{12pt} $(\mathfrak{a}_2)_1 $&	$-\frac{1}{3}$& $2$ & $1+8 q+17 q^2+46 q^3+98 q^4+O\left(q^5\right)$\\
		\rule{0pt}{12pt}$(\mathfrak{e}_6)_1$& $ -\frac{2}{3}$& $6$& $1+78 q+729 q^2+4382 q^3+19917 q^4+O\left(q^5\right)$\\
		\hline
		\hline
		\multicolumn{4}{|c|}{$\mathfrak{a}_2\simeq(A_1,D_4)$} \\
		\hline 
		VOA &$\alpha$ & $c_{\text{2d}}$ & $q$-series \\
		\hline
		\rule{0pt}{12pt}$(\mathfrak{d}_4)_1$& $-\frac{1}{2}$ & $4$ &$1+28 q+134 q^2+568 q^3+1809 q^4+O\left(q^5\right)$\\
		\hline
	   \end{tabular}
	   \caption{$SU(2)$ $N_f=4$ series}
	   \label{tab:su2}
	   \end{table}

   \subsection*{Flavored checks}

   We can also perform checks on this conjecture by further refining by the flavor symmetry. $\mathfrak{a}_0$ theory does not have flavor symmetry. $\mathfrak{a}_2$ has $SU(2)$ flavor symmetry and $\mathfrak{a}_3$ has a $SU(3)$ flavor symmetry. The integral with all the $SO(8)$ flavor fugacities can be performed. Here we specialize the fugacities for the respective Argyres-Douglas theory. The integrals for $\mathfrak{a}_0$, $\mathfrak{a}_1$ and $\mathfrak{a}_2$ with their respective flavor fugacities takes the form 
   
      \begin{equation}
	\hat{\mathcal{Z}}_{\mathfrak{g}}(q,\alpha_{\mathfrak{g}})=\frac{(q;q)^{2(1-\alpha_{\mathfrak{g}}\frac{h^\vee}{6})}}{\mathtt{N}\left(\alpha_{\mathfrak{g}}\frac{h^\vee}{6}\right)}\oint \frac{dz}{2\pi iz}\Delta(z)^{\alpha_{\mathfrak{g}}\frac{h^\vee}{6}}\text{PE}\left[\alpha_{\mathfrak{g}}\frac{h^\vee}{6}\left(\frac{-2q}{1-q}\chi_\mathbf{3}(z)+\frac{q^{\frac{1}{2}}}{1-q}\chi_{\mathbf{2}}(z)\boldsymbol{\chi}_{\mathfrak{g}}\right)\right]~,
   \end{equation}
   where $\boldmath{\chi}_{\mathfrak{g}}$ denotes the character in terms of fugacities which reproduce the correct flavored index \footnote{The embedding of the flavor fugacities in $SO(8)$ can be inferred from the Lagrangians in \cite{Maruyoshi:2016tqk, Maruyoshi:2016aim, Agarwal:2016pjo}.}
   \begin{equation}
   	\boldsymbol{\chi}_{\mathfrak{a}_0}=8,~~~\boldsymbol{\chi}_{\mathfrak{a}_1}=4\left(a+\frac{1}{a}\right),~~~
   	\boldsymbol{\chi}_{\mathfrak{a}_2}=\left(a+\frac{1}{a}+3b+\frac{3}{b}\right)~.
   \end{equation}
Since, $\mathfrak{a}_0$ has no flavor and the match for unflavored indices has been done, we only note the flavored generalized Schur limit for $\mathfrak{a}_1$ and $\mathfrak{a}_2$
\begin{equation}
	\begin{split}
	&\hat{\mathcal{Z}}_{SU(2)}(q,\alpha)=\hat{\mathcal{Z}}_{\mathfrak{a}_1}(q,3\alpha)\\&=1+\left(\frac{12 a^2 \alpha ^2+\frac{12 \alpha ^2}{a^2}+36 \alpha ^2-2 \alpha -2}{\alpha +1}\right)q\\
	&+\frac{1}{(\alpha +1) (\alpha +2)}\left(\left(a^4+\frac{1}{a^4}\right) \left(72 \alpha ^4+42 \alpha ^3+6 \alpha ^2\right)+792 \alpha ^4+18 \alpha ^3 \right. \\& \left.+\left(a^2+\frac{1}{a^2}\right) \left(432 \alpha ^4+24 \alpha ^3\right)+17 \alpha ^2-3 \alpha -2\right)q^2+O(q^3)~.
	\end{split}
\end{equation}
\begin{equation}
	\begin{split}
	&\hat{\mathcal{Z}}_{SU(2)}(q,\alpha)=\hat{\mathcal{Z}}_{\mathfrak{a}_2}(q,2\alpha)\\&=1+\left(\frac{6 a^2 \alpha ^2+\frac{6 \alpha ^2}{a^2}+24 \alpha ^2+6 a \alpha ^2 b+\frac{6 \alpha ^2 b}{a}+\frac{6 a \alpha ^2}{b}+\frac{6 \alpha ^2}{a b}-2 \alpha -2}{\alpha +1}\right)q\\&+\frac{\mathfrak{F}(\alpha,a,b)q^2}{(\alpha+1)(\alpha+2)}+O(q)^3~.
	\end{split}
\end{equation}
The expression for order $q^2$ term for $\hat{\mathcal{Z}}_{\mathfrak{a}_2}(q,\alpha_{\mathfrak{g}})$ is quite large and we denote its non-singular part in $\alpha$ by $\mathfrak{F}(\alpha,a,b)$. $\mathfrak{F}(\alpha,a,b)$ denotes a polynomial in $\alpha$ with both positive and negative powers of $a$ and $b$. It is straightforward to check that the following higher powers of monodromy traces, with flavor fugacities turned on, match the above expressions. Here we note down the expressions for traces and terms upto order $q^2$.

 \begin{equation}
 	\begin{split}
 	&(q;q)^{2}\Tr M_{\mathfrak{a}_1}(q)\\&=(q;q)^{2}\sum_{n_j,m_j\geq 0}a^{2 (n_1-m_1)}\frac{q^{\frac{\sum_{i=1}^3 \left( n_i^2+m_i^2\right)}{2}-m_2(n_1 +n_3)}}{\prod_{i=1}^3(q;q)_{n_i}(q;q)_{m_i}}\delta_{n_2-m_2,0}\delta_{n_1-m_1+n_3-m_3,0}\\
 	&=1+\left(4+\frac{2}{a^2}+2 a^2\right) q+\left(4 a^2+\frac{4}{a^2}+9\right)q^2+O(q^3)~.
 	\end{split}
 \end{equation}

 \begin{equation}
	\begin{split}
		&(q;q)^{2}\Tr M_{\mathfrak{a}_1}(q)^2\\&=(q;q)^{2}\sum_{n_j,m_j,k_j,l_j\geq 0}a^{2 (n_1-m_1+k_1-l_1)}\frac{q^{\frac{\sum_{i=1}^3 \left( n_i^2+m_i^2+k_i^2+l_i^2\right)}{2}-m_2(n_1 +n_3)-l_2(k_1+k_3)+(k_2-l_2)(n_1-m_1+n_3-m_3)}}{\prod_{i=1}^3(q;q)_{n_i}(q;q)_{m_i}(q;q)_{k_i}(q;q)_{l_i}}\\
		&\hspace{1.2in}\times \delta_{n_2-m_2+k_2-l_2,0}\delta_{n_1 - m_1 + n_3 - m_3 + k_1 - l_1 + k_3 - l_3,0}\\
		&=1+\left(46+\frac{16}{a^2}+16 a^2\right) q+\left(10 \left(a^4+\frac{1}{a^4}\right)+176 \left(a^2+\frac{1}{a^2}\right)+357\right)q^2+O(q^3
		)~.
	\end{split}
\end{equation}  

 \begin{equation}
	\begin{split}
		&(q;q)^{2}\Tr M_{\mathfrak{a}_2}(q)\\&=(q;q)^{2}\sum_{n_j,m_j\geq 0}b^{2 (n_1-m_1)} (a b)^{n_3-m_3}\frac{q^{\frac{\sum_{i=1}^4 \left( n_i^2+m_i^2\right)}{2}-m_2(n_1 +n_3+n_4)}}{\prod_{i=1}^4(q;q)_{n_i}(q;q)_{m_i}}\delta_{n_2-m_2,0}\delta_{n_1-m_1+n_3-m_3+n_4-m_4,0}\\
		&=1+\left(10+\frac{3}{b^2}+\frac{3}{a b}+\frac{3 a}{b}+\frac{3 b}{a}+3 a b+3
		b^2\right) q+a^2+\frac{1}{a^2}\\&\;\;\;\;+\left(a+\frac{1}{a}\right) \left(b^3+\frac{1}{b^3}+15 b+\frac{15}{b}\right)+15 \left(b^2+\frac{1}{b^2}\right)q^2+38+O(q^3)~.
	\end{split}
\end{equation}

	\section{Higher rank cases}
	\label{sec:higherrank}
	In this section we consider the generalized Schur limit of $USp(2N)$ $N_f=2N+2$ and $SU(N)$ $N_f=2N$. For certain positive rational values of $\alpha$, the limit coincides with Schur index of $(A_1,G)$ theories with $G$ depending on the the SCQD. Similar to the case of Argyres-Douglas theories in Section \ref{sec:su2nf4}, the generalized Schur limit is related to that of the SQCD by an overall rescaling of $\alpha$.  Here also we observe that traces of $M(q)$ coincide with the generalized Schur limit. In the lower rank Argyes-Douglas theories, the power of $M(q)$ at which the trace seems to diverge is relatively high. For the case of $(A_1,A_2)$, we were able to consider upto $\Tr M(q)^4$. In higher rank theories considered in this section, except the case of $(A_1,A_4)$, where we can consider $\Tr M(q)^2$, we only check $\Tr M(q)$. Although, due to the computational complexity for higher ranks,  we have not verified that considering higher powers of $M(q)$ may not lead to a convergent $q$-series, from the poles in the expression of the generalized Schur limit, it appears that some sort of divergence might appear . We also check the trace refined by the flavor fugacities at very low orders. For performing a flavored check, we use the convention for the fugacities such that the $(q;q)^{2\mathfrak{r}}\Tr M(q)^{-1}$ equals the Schur index. The flavored traces can be computed similar to those in Section \ref{sec:su2nf4} and since the explicit expressions for traces for the theories discussed in this section are lengthy, we do not display them here.  Let us note that as in the case of $SU(2)$ SQCD, this limit of the SQCD index may be equal to $q$-series characters which may not be related to the trace of $M(q)$ for some Argyres-Douglas theory. 
	
	\subsection{$USp(2N)$ $N_f=2N+2$ series}

\begin{table}[t]
	\centering
		\begin{tabular}{|c|c|c|c|c|c|}
			\hline
			Theory & $12c$ & $24a$  & $\Delta_{i}$ &$\alpha$\\
			\hline
			$USp(2N)$+$(2N+2)$F &$2 N (4 N+3)$ &$N(14 N+9)$& $(2i-1)+1$ & $1$\\
			\hline
			$D_2(SU(2N+1))$ & $4N(N+1)$ & $7N(N+1)$ & $\frac{(2i-1)}{2}+1$ & $\frac{1}{2}$\\
			\hline
			$(A_1,D_{2N+1})$ &$6N$ &$\frac{3 N (8 N+3)}{2 N+1}$ & $\frac{(2i-1)}{2N+1}+1$ & $\frac{1}{2N+1}$\\
			\hline
			$(A_1,A_{2N})$ &$\frac{2 N (6 N+5)}{2 N+3}$ & $\frac{n (24 N+19)}{2 N+3}$& $\frac{(2i-1)}{2N+3}+1$ &$\frac{1}{2N+3}$\\
			\hline
	\end{tabular}
	\caption{	Series obtained from $USp(2N)$ $N_f=2N+2$. }
	\label{tab:usp2nalphapos}
\end{table}
	The generalized Schur limit for $USp(2N)$ $N_f=2N+2$ for particular rational values of alpha is displayed in Table \ref{tab:usp2nalphapos}. The 2d central charge for this limit takes the form
\begin{equation}
	c_{\text{2d}}=\alpha  \left(-8 N^2-4 N\right)-2 N~.
\end{equation}
 This limit for the $(A_1,D_{2N+1})$ theory is related to the $USp(2N)$ by the rescaling of $\alpha$ by $2N+1$ and for $(A_1,A_{2N})$ is related by rescaling $\alpha$ by $2N+3$. The generalized Schur limit for $USp(4)$ solves a third order MLDE and for $USp(6)$ solves a fourth order MLDE. The coefficients for these MLDEs are collected in Appendix \ref{app:notations}. Using this MLDE data, the limit can be evaluated and the first few terms in the $q$-series expansion are as follows
		\begin{equation}
			\begin{split}
				\hat{\mathcal{Z}}_{USp(4)}(q,\alpha)&=1+\frac{4 \left(-1-3 \alpha +70 \alpha ^2\right) q}{1+3 \alpha }\\ &+\frac{\left(4+22 \alpha -944 \alpha ^2-2782 \alpha
					^3+52500 \alpha ^4+39200 \alpha ^5\right) q^2}{2+11 \alpha +18 \alpha ^2+9 \alpha ^3}\\&+\frac{8 \left(18+117 \alpha
					-2189 \alpha ^2-9557 \alpha ^3+35011 \alpha ^4\right)q^3}{9
					(1+\alpha )^2 \left(2+9 \alpha +9 \alpha ^2\right)}\\&+\frac{8\left(27580 \alpha ^5+2944900 \alpha ^6+1372000 \alpha ^7\right) q^3}{9
					(1+\alpha )^2 \left(2+9 \alpha +9 \alpha ^2\right)}+O\left(q^4\right)~.
			\end{split}
		\end{equation}
		
		\begin{equation}
			\begin{split}
				\hat{\mathcal{Z}}_{USp(6)}(q,\alpha)&=1+\frac{6 \left(126 \alpha ^2-5 \alpha -1\right) q}{5 \alpha +1}\\&+\frac{9 \left(95256 \alpha ^5+44310 \alpha ^4-3789 \alpha ^3-560 \alpha ^2+21 \alpha +2\right) q^2}{(5 \alpha +1) \left(15 \alpha ^2+11 \alpha +2\right)}\\&+\frac{2 \left(108020304 \alpha ^8+199608948 \alpha ^7+94489416 \alpha ^6-6647523 \alpha ^5\right)q^3}{(5 \alpha +1) \left(5 \alpha ^2+8 \alpha +3\right) \left(15 \alpha ^2+11 \alpha +2\right)}\\ &+\frac{2\left(-911900 \alpha ^4-13820 \alpha ^3-1210 \alpha ^2+395 \alpha +30\right) q^3}{(5 \alpha +1) \left(5 \alpha ^2+8 \alpha +3\right) \left(15 \alpha ^2+11 \alpha +2\right)}+O\left(q^4\right)~.
			\end{split}
		\end{equation}

	The values of $\alpha$ for which we find integer coefficient $q$-series are collected in table \ref{tab:usp4} and \ref{tab:usp6}. For the $USp(4)$ case, we find that for $\alpha=-\frac{1}{7}, -\frac{2}{7}$ correspond to the $q$-series expansion of $\mathfrak{osp}(1|4)_1$ and $(X_1)_1$ VOA. $\alpha=-\frac{1}{5}$ corresponds to the $q$-series expansion of $\left(\mathfrak{a}_4\right)_1$. As discussed in previous sections, by an overall rescaling of $\alpha$, these results can be interpreted as generalized Schur limits of the Argyres-Douglas theories and are equal to the trace of higher powers of the quantum monodromy considered in \cite{Kim:2024dxu}, which is consistent with our conjecture. We refer the reader to \cite{Kim:2024dxu} for the computation of these traces.  The $USp(6)$ case for $\alpha=-\frac{1}{7},-\frac{1}{9}$ correspond to the characters of $(\mathfrak{a}_6)_1$ and $\mathfrak{osp}(1|6)$ \cite{Creutzig:2024ljv}. We perform the flavored checks below.

	\begin{table}[t]
		\centering
		\begin{tabular}{|c|c|c|c|}
			\hline
			\multicolumn{4}{|c|}{$(A_1,A_4)$}\\
			\hline
			VOA&$\alpha$ & $c_{\text{2d}}$ & $q$-series \\
			\hline
				\rule{0pt}{12pt} $\mathfrak{osp}(1|4)_1 $& $ -\frac{1}{7}$& $\frac{12}{7}$&$1+6 q+12 q^2+28 q^3+O\left(q^4\right)$\\
			\rule{0pt}{12pt} $(X_1)_1 $& $-\frac{2}{7}$&$\frac{52}{7}$ &$1+156 q+2236 q^2+17056 q^3+O\left(q^4\right)$\\
			\hline
			\hline
			\multicolumn{4}{|c|}{$(A_1,D_5)$} \\
			\hline 
			VOA&$\alpha$ & $c_{\text{2d}}$ & $q$-series \\
			\hline
			\rule{0pt}{12pt} $(\mathfrak{a}_4)_1$ & 	$-\frac{1}{5}$&$4$ &$1+24 q+124 q^2+500 q^3+O\left(q^4\right)$\\
			\hline
		\end{tabular}
		\caption{$USp(4)$ $N_f=6$ series} 
		\label{tab:usp4}
	\end{table}

	\begin{table}[t]
		\centering
		\begin{tabular}{|c|c|c|c|}
			\hline
			\multicolumn{4}{|c|}{$(A_1,A_6)$} \\
			\hline
			VOA&$\alpha$ & $c_{\text{2d}}$ & $q$-series \\
			\hline
		\rule{0pt}{12pt} 	$\mathfrak{osp}(1|6)_1$&	$-\frac{1}{9}$&$\frac{10}{3}$ &$1+15 q+65 q^2+220 q^3+O\left(q^4\right)$\\
				\hline
				\hline
			\multicolumn{4}{|c|}{$(A_1,D_7)$ }\\
			\hline
			&$\alpha$ & $c_{\text{2d}}$ & $q$-series \\
				\hline
		\rule{0pt}{12pt} 	$	\left(\mathfrak{a}_6\right)_{1}$	& $ -\frac{1}{7}$& $6$ &$1+48 q+489 q^2+2842 q^3+O\left(q^4\right)$\\
			\hline
		\end{tabular}
		\caption{$USp(6)$ $N_f=8$ series}
		\label{tab:usp6}
	\end{table}
	
     \subsection*{Flavored checks}
    The BPS quivers for $(A_1,A_{2N})$ and $(A_1,D_{2N+1})$ are given in Figure \ref{fig:bpsquivad} and the flavor symmetry and conventions for the trace of flavor vector are collected in Table \ref{tab:flavconv}. $(A_1,A_{2N})$ has no flavor symmetry so we only do flavored computations for $(A_1,D_{2N+1})$ case. Upto $O(q)^2$, the $q$-series for $(A_1,D_5)$ and $(A_1,D_7)$, and the trace of $M(q)$ take the following form

	\subsubsection*{\underline{$(A_1,D_5)$}}
	\begin{equation}
	\begin{split}
		&\hat{\mathcal{Z}}_{USp(4)}(q,\alpha)=\hat{\mathcal{Z}}_{(A_1,D_5)}(q,5\alpha)=\\=&1+\left(\frac{40 \alpha ^2(a^2+\frac{1}{a^2})+200 \alpha ^2-12 \alpha -4}{3 \alpha +1}\right)q\\
		&+\frac{2q^2}{(\alpha +1) (3 \alpha +1) (3 \alpha +2)}\left( \left(a^4+\frac{1}{a^4}\right) \left(400 \alpha ^5+530 \alpha ^4+140 \alpha ^3+10 \alpha ^2\right)\right.\\&+\left. \left(10800 \alpha ^5+14150 \alpha ^4-1191 \alpha ^3-332 \alpha ^2+11 \alpha +2\right)\right.\\&+\left.  \left(a^2+\frac{1}{a^2}\right) \left(4000 \alpha ^5+5520 \alpha ^4-240 \alpha ^3-80 \alpha ^2\right)\right)+O(q^3)~.
	\end{split}
\end{equation}
	
		 \begin{equation}
			(q;q)^{4}\Tr M_{(A_1,D_5)}(q)=1+\left(16+\frac{4}{a^2}+4 a^2\right) q+\left(68+\frac{28}{a^2}+28
			a^2\right) q^2+O(q^3)~.
	\end{equation}
	
	\subsubsection*{\underline{$(A_1,D_7)$}}
	\begin{equation}
	\begin{split}
		&\hat{\mathcal{Z}}_{USp(6)}(q,\alpha)=\hat{\mathcal{Z}}_{(A_1,D_7)}(q,7\alpha)\\=&1+\left(\frac{84 \left(a+\frac{1}{a}+7\right) \alpha ^2-30 \alpha -6}{5 \alpha +1}\right)q\\
		&+\frac{q^2}{(3 \alpha +1) (5 \alpha +1) (5 \alpha +2)}\left(539784 \alpha ^5+240786 \alpha ^4-27549 \alpha ^3-3780 \alpha ^2+189 \alpha +18\right.\\&+\left. \left(a^4+\frac{1}{a^4}\right) \left(10584 \alpha ^5+5922 \alpha ^4+924 \alpha ^3+42 \alpha ^2\right)\right.\\&+\left.\left(a^2+\frac{1}{a^2}\right) \left(148176 \alpha ^5+73080 \alpha ^4-4200 \alpha ^3-672 \alpha ^2\right)\right)+O(q^3)~.
	\end{split}
\end{equation}

	 \begin{equation}
		(q;q)^{6}\Tr M_{(A_1,D_7)}(q)=1+\left(36+\frac{6}{a^2}+6 a^2\right) q+\left(297+\frac{96}{a^2}+96
			a^2\right) q^2+O(q^3)~.
	\end{equation}
	
	\subsection{$SU(N)$ $N_f=2N$}
	
	\begin{table}[t]
		\centering
			\begin{tabular}{|c|c|c|c|c|c|}
				\hline
				Theory & $12c$ & $24a$  & $\Delta_{i}$ & $\alpha$\\
				\hline
				$R_{2,2N-1}$ &$2\left(4 N^2-N-1\right)$ &$14 N^2-5 N-5$  & $2i+1$ &$2$\\
				\hline
				$SU(N)$+$2N$F &$4 N^2-2$ &$7 N^2-5$& $i+1$ & $1$\\
				\hline
				$(A_1,D_{2N})$ &$6N-4$ &$2 (6 N-5)$ & $\frac{i}{N}+1$ & $\frac{1}{N}$\\
				\hline
				$(A_1,A_{2N-1})$ &$\frac{6 N^2-2 N-2}{N+1}$ & $\frac{12 N^2-5 N-5}{N+1}$& $\frac{i}{N+1}+1$  & $\frac{1}{N+1}$\\
				\hline
		\end{tabular}
		\caption{Series obtained from $SU(N)$ $N_f=2N$.}
		\label{tab:sunalphapos}
	\end{table}
	The generalized Schur limit for for $USp(2N)$ $N_f=2N+2$ for particular rational values of $\alpha$ is displayed in Table \ref{tab:sunalphapos}. The 2d central charge for this limit takes the form
	\begin{equation}
		c_{\text{2d}}=\alpha  \left(2 N-4 N^2\right)-2 N+2~.
	\end{equation}
	This limit for the $(A_1,D_{2N})$ theory is related to the $SU(N)$ by the rescaling of $\alpha$ by $N$ and for $(A_1,A_{2N-1})$ is related by rescaling $\alpha$ by $N+1$.  The generalized Schur limit for $SU(3)$ solves a fourth order twisted MLDE and for $SU(4)$ solves a sixth order MLDE.  Note that indices for $SU(N)$ for odd $N$ have half integer powers in $q$ and for even $N$ have integer powers in $q$. The coefficients for these MLDEs are collected in Appendix \ref{app:notations}. The first few terms in the $q$-series expansion of the generalized Schur limit for the $SU(3)$ theory with $6$ flavors and $SU(4)$ theory with $8$ flavors are as follows 
	\begin{equation}
		\hat{\mathcal{Z}}_{SU(3)}(q,\alpha)=1+\left(\frac{120 \alpha ^2-8 \alpha -4}{2 \alpha +1}\right)q+\frac{240 \alpha ^3}{(\alpha +1) (2 \alpha +1)}q^{3/2}+O(q^2)~.
	\end{equation}

	\begin{equation}
		\begin{split}
		\hat{\mathcal{Z}}_{SU(4)}(q,\alpha)&=1+\left(\frac{280 \alpha ^2-18 \alpha -6}{3 \alpha +1}\right)q\\&+\left(\frac{78400 \alpha ^6+144760 \alpha ^5+54902 \alpha ^4-9955 \alpha ^3-1740 \alpha ^2+135 \alpha +18}{(\alpha +1) (2 \alpha +1) (3 \alpha +1) (3 \alpha +2)}\right)q^2+O(q^3)~.
		\end{split}
	\end{equation}	
The values of $\alpha$ for which we find integer coefficient $q$-series matching the trace monodromies are collected in table \ref{tab:su3} and \ref{tab:su4}. For theories in this section, we have not been able to identify the VOAs corresponding to the $q$-series.
	\begin{table}[t]
		\centering
		\begin{tabular}{|c|c|c|}
			\hline
			\multicolumn{3}{|c|}{$(A_1,A_5)$}\\
				\hline
			$\alpha$ & $c_{\text{2d}}$ & $q$-series \\
			\hline
			\rule{0pt}{12pt} $-\frac{1}{4}$& $\frac{7}{2}$ &$1+11 q-10 q^{3/2}+42 q^2-40 q^{5/2}+138 q^3-162 q^{7/2}+O\left(q^4\right)$\\
			\hline
			\hline
			\multicolumn{3}{|c|}{$(A_1,D_6)$}\\
			\hline
			$\alpha$ & $c_{\text{2d}}$ & $q$-series \\
			\hline
			\rule{0pt}{12pt} $-\frac{1}{3}$& $6$ &$1+36q-40q^{3/2}+297q^2-360q^{5/2}+1590q^3-2160q^{7/2}+O(q^4)$\\
			\hline
		\end{tabular}
		\caption{$SU(3)$ $N_f=6$ series}
		\label{tab:su3} 
	\end{table}

		\begin{table}[t]
		\centering
		\begin{tabular}{|c|c|c|}
				\hline
				\multicolumn{3}{|c|}{$(A_1,D_8)$} \\
				\hline
				\rule{0pt}{12pt} 	$\alpha$ & $c_{\text{2d}}$ & $q$-series \\
				\hline
					\rule{0pt}{12pt} $-\frac{1}{4}$& $8$ &$1+64q+1052q^2+O(q^3)$\\
				\hline
				\hline
				\multicolumn{3}{|c|}{$(A_1,A_7)$}\\
				\hline
				\rule{0pt}{12pt} 	$\alpha$ & $c_{\text{2d}}$ & $q$-series \\
				\hline
				\rule{0pt}{12pt} 	$-\frac{1}{5}$& $\frac{26}{5}$ &$1+22q+177q^2+O(q^3)$\\
				\hline
		\end{tabular}
		\caption{$SU(4)$ $N_f=8$ series}
		\label{tab:su4}
	\end{table}

      \subsection*{Flavored checks}
       The BPS quivers for $(A_1,A_{2N-1})$ and $(A_1,D_{2N})$ are given in Figure \ref{fig:bpsquivad} and the flavor symmetry and conventions for the trace of flavor vector are collected in Table \ref{tab:flavconv}.   We perform flavored checks for this case as well. Upto $O(q)^2$, the $q$-series takes the following form
    
    \subsubsection*{\underline{$(A_1,A_5)$}}
    \begin{equation}
    	\hat{\mathcal{Z}}_{SU(3)}(q,\alpha)=\hat{\mathcal{Z}}_{(A_1,A_5)}(q,4\alpha)=1+\left(\frac{120 \alpha ^2-8 \alpha -4}{2 \alpha +1}\right)q+\left(\frac{120\alpha^3\left(a+\frac{1}{a}\right)}{(\alpha +1) (2 \alpha +1)}\right)q^{3/2}+O(q^2)~.
    \end{equation}
    
    \begin{equation}
    	(q;q)^{4}\Tr M_{(A_1,A_5)}(q)=1+11 q-5\left(a+\frac{1}{a} \right) q^{3/2}+O(q^2)~.
    \end{equation}
    \subsubsection*{\underline{$(A_1,D_6)$}}
    \begin{equation}
    \begin{split}
    &\hat{\mathcal{Z}}_{SU(3)}(q,\alpha)=\hat{\mathcal{Z}}_{(A_1,D_6)}(q,3\alpha)\\&=1+\left(\frac{90 \alpha ^2-8 \alpha +15 \alpha ^2 b^2+\frac{15 \alpha ^2}{b^2}-4}{2 \alpha +1}\right)q+\left(\frac{60\alpha^3\left(a+\frac{1}{a}\right)\left(b+\frac{1}{b}\right)}{(\alpha +1) (2 \alpha +1)}\right)q^{3/2}+O(q^2)~.
    \end{split}
    \end{equation}
    
    \begin{equation}
    	(q;q)^{4}\Tr M_{(A_1,D_6)}(q)=1+q \left(5 b^2+\frac{5}{b^2}+26\right)-10  \left(a+\frac{1}{a}\right) \left(b+\frac{1}{b}\right)q^{3/2}+O\left(q^2\right)~.
    \end{equation}
    
    \subsubsection*{\underline{$(A_1,A_7)$}}
    \begin{equation}
    \begin{split}
    &\hat{\mathcal{Z}}_{SU(4)}(q,\alpha)=\hat{\mathcal{Z}}_{(A_1,A_7)}(q,5\alpha)\\
    &=1+\left(\frac{280 \alpha ^2-18 \alpha -6}{3 \alpha +1}\right)q\\&+\frac{q^2}{(\alpha +1) (2 \alpha +1) (3 \alpha +1) (3 \alpha +2)}\left(2 \left(a+\frac{1}{a}\right) \alpha ^4 (2520 \alpha +1680)\right.\\&\left.+\left(78400 \alpha ^6+134680 \alpha ^5+48182 \alpha ^4-9955 \alpha ^3-1740 \alpha ^2+135 \alpha +18\right)\right)+O(q^3)~.
    \end{split}
    \end{equation}
    
    \begin{equation}
    	(q;q)^{6}\Tr M_{(A_1,A_7)}(q)=1+22 q+q^2 \left(14 a+\frac{14}{a}+149\right)+O\left(q^3\right)~.
    \end{equation}

    \subsubsection*{\underline{$(A_1,D_8)$}}
    \begin{equation}
    \begin{split}
    &\hat{\mathcal{Z}}_{SU(4)}(q,\alpha)=\hat{\mathcal{Z}}_{(A_1,D_8)}(q,4\alpha)\\
    &=
    1+\left(\frac{-18 \alpha +\alpha ^2 \left(28 b^2+\frac{28}{b^2}+224\right)-6}{3 \alpha +1}\right)q+\frac{\mathfrak{G}(a,b,\alpha)~q^2}{(\alpha +1) (2 \alpha +1) (3 \alpha +1) (3 \alpha +2)}+O(q^3)~.
    \end{split}
    \end{equation}
    The expresion for $\mathfrak{G}(a,b,\alpha)$ is too long so we do not write it explicitly. $\mathfrak{G}(a,b,\alpha)$ is a polynomial in $\alpha$ and contains positive and negative powers of $a$ and $b$.
        \begin{equation}
        	\begin{split}
    	(q;q)^{6}\Tr M_{(A_1,D_8)}(q)&=1+\left(7 b^2+\frac{7}{b^2}+50\right)q\\&+\left(35 \left( a b^2+\frac{1}{a b^2}\right)+35 \left(a+\frac{1}{a}\right)+161\left( b^2+\frac{1}{b^2}\right)+590\right)q^2+O(q^3)~.
    	\end{split}
    \end{equation}

	\section*{Acknowldegements}
	We would like to thank Leonardo Rastelli and Palash Singh for helpful discussions. AD is grateful to Brandon C. Rayhaun for discussions regarding identification of $q$-series to VOA characters and  to Shlomo S. Razamat for comments on the draft. The work of AD is supported in part by NSF grant PHY-2513893 and by the
	Simons Foundation grant 681267 (Simons Investigator Award).
	\appendix
	\section{Notations and MLDE coefficients}
\label{app:notations}
We follow the notations in \cite{Beem:2017ooy} (see also \cite{Beem:2021zvt}). Here we collect only relevant formulae and notation.	A monic modular linear differential operator (MLDO) is defined as
\begin{equation}
	\mathcal{D}^{(k)}_q:=D^{(k)}_q+\sum_{r=1}^k f_r(q) D_q^{k-r},~~f_r(q)\in M_{2r}(\widetilde{\Gamma},\mathbb{C})~,
\end{equation}
where $M_{k}(\widetilde{\Gamma},\mathbb{C})$ denotes a modular form of weight $k$ and transformations in some congruence subgroup $\widetilde{\Gamma}$ of the full modular group $\Gamma$. $D_q^{(k)}$ denote derivatives composed of Ramanujan–Serre derivative $\partial_{(k)}$
\begin{equation}
	D_q^{(k)}:=\partial_{2k-2}\dots\partial_{2}\partial_{0},~~\partial_{(k)}:=q\partial_q+k\mathbb{E}_2(\tau)~.
\end{equation}
The convention for Eisenstein series in this paper is ($q=e^{2\pi i\tau}$)
\begin{equation}
	\mathbb{E}_{2k}(\tau):=-\frac{B_{2k}}{(2k)!}+\frac{2}{(2k-1)!}\sum_{n\geq 1}\frac{n^{2k-1}q^n}{1-q^n}~,
\end{equation}
where $B_{2k}$ is the $2k$-th Bernoulli number. For Schur indices with integer powers of $q $, the modular group under consideration is the full modular group and the MLDO takes the form
\begin{equation}
	\label{eq:untwist}
	D^{(k)}_q+\sum_{n=1}^k\sum_{i,j}\lambda_{2n,i,j}\mathbb{E}_{4}(\tau)^i\mathbb{E}_{6}(\tau)^j D_q^{(k-n)}~,
\end{equation}
where $i,j$ are summed such that $\mathbb{E}_{4}(\tau)^i\mathbb{E}_{6}(\tau)^j$ is weight $2n$ modular form. For Schur indices with half integer powers of $q$, the relevant congruence subgroup is $\Gamma^{0}(2)$ and MLDO takes the form

\begin{equation}
	\label{eq:twist}
	D^{(k)}_q+\sum_{n=1}^k\sum_{i=0}^{\left[\frac{n}{2}\right]}\lambda_{2n,i}\Theta_{i,n-i}(\tau) D_q^{(k-n)}
\end{equation}
where
\begin{equation}
	\Theta_{r,s}(\tau):=\theta_2(\tau)^{4r}\theta_3(\tau)^{4s}+\theta_2(\tau)^{4s}\theta_3(\tau)^{4r},~~r\leq s
\end{equation}
and
\begin{equation}
	\theta_2(\tau)=\sum_{n=-\infty}^\infty q^{\frac{1}{2}\left(n+\frac{1}{2}\right)^2},~~\theta_3(\tau)=\sum_{n=-\infty}^\infty q^{\frac{n^2}{2}}~.
\end{equation}

\subsection*{MLDE coefficients}
Here, we collect the MLDE coefficients for the examples discussed in the main text. The coefficients $\lambda_{2n,i,j}$ are the coefficients for untwisted MLDEs and $\lambda_{2n,i}$ are for twisted MLDEs as described in equation \eqref{eq:untwist} and \eqref{eq:twist}. \newline\newline
	\underline{$USp(4)$ $N_f=6$}
	\begin{equation}
		\begin{split}
		\lambda _{4,1,0}&= 20-1680 \alpha ^2,\\\lambda _{6,0,1}&= 22400 \alpha ^3-11760 \alpha ^2+140~.
		\end{split}
	\end{equation}
	\underline{$USp(6)$ $N_f=8$}
	\begin{equation}
		\begin{split}
			\lambda _{4,1,0}&= 50-7560 \alpha ^2,\\ \lambda _{6,0,1}&= 241920 \alpha ^3-105840 \alpha ^2+700,\\\lambda _{8,2,0}&= 3402000 \alpha ^4+1036800 \alpha ^3-340200 \alpha ^2+2025~.
		\end{split}
	\end{equation}
	\underline{$SU(3)$ $N_f=6$}
    \begin{equation}
    	\begin{split}
    		\lambda _{1,0}&= 0.333333\, -0.5 \alpha ,\\ \lambda _{2,0}&= 0.0277778\, -1. \alpha ^2,\\ \lambda _{2,1}&= 0.5 \alpha ^2-0.0138889,\\ \lambda _{3,0}&= 0.03125 \alpha ^3+0.166667 \alpha ^2-0.0138889 \alpha -0.00462963,\\ \lambda _{3,1}&= 0.46875 \alpha ^3-0.75 \alpha ^2+0.0208333,\\ \lambda _{4,0}&= 0.0585938 \alpha ^4-0.00520833 \alpha ^3-0.0277778 \alpha ^2+0.00231481 \alpha +0.000771605,\\ \lambda _{4,1}&= 0.703125 \alpha ^4-0.427083 \alpha ^3+0.138889 \alpha ^2-0.00115741 \alpha -0.00385802,\\ \lambda _{4,2}&= -0.761719 \alpha ^4+0.515625 \alpha ^3-0.208333 \alpha ^2-0.00347222 \alpha +0.00578704~.
    	\end{split}
    \end{equation}
    \underline{$SU(4)$ $N_f=8$}
    \begin{equation}
    \begin{split}
    \lambda _{4,1,0}= & -4800 \alpha ^2+1200 \alpha -125~,
    \\ \lambda _{6,0,1} = & 56000 \alpha ^3-117600 \alpha ^2+2100~,\\
    \lambda _{8,2,0}= &  5.2416\times 10^6 \alpha ^4-2.16\times 10^6 \alpha ^3-744000\alpha ^2+60000\alpha +11275~,\\
    \lambda _{10,1,1}=& -6.1824\times 10^7 \alpha ^5+1.2553\times 10^8 \alpha ^4-2.996\times 10^7 \alpha ^3-1.1004\times 10^7 \alpha ^2\\&+840000 \alpha +158900~,\\
    \lambda _{12,0,2} = & -0.000151088 \alpha ^8+0.00220808 \alpha ^7-1.75616\times 10^9 \alpha ^6+1.0537\times 10^9 \alpha ^5\\&-1.18541\times 10^8 \alpha ^4+9.408\times 10^7 \alpha ^3-4.9392\times 10^7 \alpha ^2+0.175168 \alpha +793800~,\\
    \lambda _{12,3,0} = & 0.0000751684 \alpha ^7-0.000462009 \alpha ^6-2.6496\times 10^8 \alpha ^5+5.07744\times 10^8 \alpha ^4\\&-1.356\times 10^8 \alpha ^3-1.224\times 10^7 \alpha ^2+2.43\times 10^6 \alpha +131625~.
    \end{split}
    \end{equation}
	\bibliographystyle{JHEP.bst}	
	\bibliography{bibliref.bib}

@article{Kim:2024dxu,
	author = "Kim, Heeyeon and Song, Jaewon",
	title = "{A Family of Vertex Operator Algebras from Argyres-Douglas Theory}",
	eprint = "2412.20015",
	archivePrefix = "arXiv",
	primaryClass = "hep-th",
	month = "12",
	year = "2024"
}

@article{Cecotti:2015lab,
	author = "Cecotti, Sergio and Song, Jaewon and Vafa, Cumrun and Yan, Wenbin",
	title = "{Superconformal Index, BPS Monodromy and Chiral Algebras}",
	eprint = "1511.01516",
	archivePrefix = "arXiv",
	primaryClass = "hep-th",
	doi = "10.1007/JHEP11(2017)013",
	journal = "JHEP",
	volume = "11",
	pages = "013",
	year = "2017"
}

@article{Deb:2025ypl,
	author = "Deb, Anirudh and Razamat, Shlomo S.",
	title = "{Generalized Schur partition functions and RG flows}",
	eprint = "2506.13764",
	archivePrefix = "arXiv",
	primaryClass = "hep-th",
	reportNumber = "YITP-SB-2025-12",
	month = "6",
	year = "2025"
}

@article{Arakawa:2016hkg,
	author = "Arakawa, Tomoyuki and Kawasetsu, Kazuya",
	title = "{Quasi-lisse vertex algebras and modular linear differential equations}",
	eprint = "1610.05865",
	archivePrefix = "arXiv",
	primaryClass = "math.QA",
	month = "10",
	year = "2016"
}

@article{Cecotti:2010fi,
	author = "Cecotti, Sergio and Neitzke, Andrew and Vafa, Cumrun",
	title = "{R-Twisting and 4d/2d Correspondences}",
	eprint = "1006.3435",
	archivePrefix = "arXiv",
	primaryClass = "hep-th",
	month = "6",
	year = "2010"
}

@article{Cordova:2015nma,
	author = "Cordova, Clay and Shao, Shu-Heng",
	title = "{Schur Indices, BPS Particles, and Argyres-Douglas Theories}",
	eprint = "1506.00265",
	archivePrefix = "arXiv",
	primaryClass = "hep-th",
	doi = "10.1007/JHEP01(2016)040",
	journal = "JHEP",
	volume = "01",
	pages = "040",
	year = "2016"
}

@article{Cordova:2016uwk,
	author = "Cordova, Clay and Gaiotto, Davide and Shao, Shu-Heng",
	title = "{Infrared Computations of Defect Schur Indices}",
	eprint = "1606.08429",
	archivePrefix = "arXiv",
	primaryClass = "hep-th",
	doi = "10.1007/JHEP11(2016)106",
	journal = "JHEP",
	volume = "11",
	pages = "106",
	year = "2016"
}

@article{Cordova:2017ohl,
	author = "Cordova, Clay and Gaiotto, Davide and Shao, Shu-Heng",
	title = "{Surface Defect Indices and 2d-4d BPS States}",
	eprint = "1703.02525",
	archivePrefix = "arXiv",
	primaryClass = "hep-th",
	doi = "10.1007/JHEP12(2017)078",
	journal = "JHEP",
	volume = "12",
	pages = "078",
	year = "2017"
}

@article{Cordova:2017mhb,
	author = "Cordova, Clay and Gaiotto, Davide and Shao, Shu-Heng",
	title = "{Surface Defects and Chiral Algebras}",
	eprint = "1704.01955",
	archivePrefix = "arXiv",
	primaryClass = "hep-th",
	doi = "10.1007/JHEP05(2017)140",
	journal = "JHEP",
	volume = "05",
	pages = "140",
	year = "2017"
}

@article{Kaidi:2022sng,
	author = "Kaidi, Justin and Martone, Mario and Rastelli, Leonardo and Weaver, Mitch",
	title = "{Needles in a haystack. An algorithmic approach to the classification of 4d $ \mathcal{N} $ = 2 SCFTs}",
	eprint = "2202.06959",
	archivePrefix = "arXiv",
	primaryClass = "hep-th",
	reportNumber = "YITP-SB-2022-02",
	doi = "10.1007/JHEP03(2022)210",
	journal = "JHEP",
	volume = "03",
	pages = "210",
	year = "2022"
}

@article{Buican:2024jjl,
	author = "Buican, Matthew",
	title = "{Coulomb branch operator algebras and universal selection rules for $ \mathcal{N} $ = 2 SCFTs}",
	eprint = "2406.00178",
	archivePrefix = "arXiv",
	primaryClass = "hep-th",
	doi = "10.1007/JHEP07(2025)212",
	journal = "JHEP",
	volume = "07",
	pages = "212",
	year = "2025"
}

@article{Fredrickson:2017yka,
	author = "Fredrickson, Laura and Pei, Du and Yan, Wenbin and Ye, Ke",
	title = "{Argyres-Douglas Theories, Chiral Algebras and Wild Hitchin Characters}",
	eprint = "1701.08782",
	archivePrefix = "arXiv",
	primaryClass = "hep-th",
	reportNumber = "CALT-TH-2016-038",
	doi = "10.1007/JHEP01(2018)150",
	journal = "JHEP",
	volume = "01",
	pages = "150",
	year = "2018"
}

@article{Pan:2024hcz,
	author = "Pan, Yiwen and Yan, Wenbin",
	title = "{Mirror symmetry for circle compactified 4d $A_1$ class-$S$ theories}",
	eprint = "2410.15695",
	archivePrefix = "arXiv",
	primaryClass = "hep-th",
	month = "10",
	year = "2024"
}

@article{Deb:2025cqr,
	author = "Deb, Anirudh and Meneghelli, Carlo and Rastelli, Leonardo",
	title = "{The Nilpotency Index for 4d $\mathcal{N}=2$ SCFTs}",
	eprint = "2503.05975",
	archivePrefix = "arXiv",
	primaryClass = "hep-th",
	reportNumber = "YITP-SB-2025-05",
	month = "3",
	year = "2025"
}

@article{Li:2025nhc,
	author = "Li, Yutong and Pan, Yiwen and Yan, Wenbin",
	title = "{Chiral algebra, Wilson lines, and mixed Hodge structure of Coulomb branch}",
	eprint = "2510.03888",
	archivePrefix = "arXiv",
	primaryClass = "hep-th",
	month = "10",
	year = "2025"
}

@article{Beem:2017ooy,
	author = "Beem, Christopher and Rastelli, Leonardo",
	title = "{Vertex operator algebras, Higgs branches, and modular differential equations}",
	eprint = "1707.07679",
	archivePrefix = "arXiv",
	primaryClass = "hep-th",
	reportNumber = "YITP-SB-17-27",
	doi = "10.1007/JHEP08(2018)114",
	journal = "JHEP",
	volume = "08",
	pages = "114",
	year = "2018"
}

@article{Argyres:1995jj,
	author = "Argyres, Philip C. and Douglas, Michael R.",
	title = "{New phenomena in SU(3) supersymmetric gauge theory}",
	eprint = "hep-th/9505062",
	archivePrefix = "arXiv",
	reportNumber = "IASSNS-HEP-95-31, RU-95-28",
	doi = "10.1016/0550-3213(95)00281-V",
	journal = "Nucl. Phys. B",
	volume = "448",
	pages = "93--126",
	year = "1995"
}

@article{Argyres:1995xn,
	author = "Argyres, Philip C. and Plesser, M. Ronen and Seiberg, Nathan and Witten, Edward",
	title = "{New N=2 superconformal field theories in four-dimensions}",
	eprint = "hep-th/9511154",
	archivePrefix = "arXiv",
	reportNumber = "RU-95-81, WIS-95-59-PH, IASSNS-HEP-95-95",
	doi = "10.1016/0550-3213(95)00671-0",
	journal = "Nucl. Phys. B",
	volume = "461",
	pages = "71--84",
	year = "1996"
}

@article{Minahan:1996fg,
	author = "Minahan, Joseph A. and Nemeschansky, Dennis",
	title = "{An N=2 superconformal fixed point with E(6) global symmetry}",
	eprint = "hep-th/9608047",
	archivePrefix = "arXiv",
	reportNumber = "USC-96-18",
	doi = "10.1016/S0550-3213(96)00552-4",
	journal = "Nucl. Phys. B",
	volume = "482",
	pages = "142--152",
	year = "1996"
}

@article{Minahan:1996cj,
	author = "Minahan, Joseph A. and Nemeschansky, Dennis",
	title = "{Superconformal fixed points with E(n) global symmetry}",
	eprint = "hep-th/9610076",
	archivePrefix = "arXiv",
	reportNumber = "USC-96-23",
	doi = "10.1016/S0550-3213(97)00039-4",
	journal = "Nucl. Phys. B",
	volume = "489",
	pages = "24--46",
	year = "1997"
}

@article{Gadde:2011uv,
	author = "Gadde, Abhijit and Rastelli, Leonardo and Razamat, Shlomo S. and Yan, Wenbin",
	title = "{Gauge Theories and Macdonald Polynomials}",
	eprint = "1110.3740",
	archivePrefix = "arXiv",
	primaryClass = "hep-th",
	reportNumber = "YITP-SB-11-30",
	doi = "10.1007/s00220-012-1607-8",
	journal = "Commun. Math. Phys.",
	volume = "319",
	pages = "147--193",
	year = "2013"
}

@article{Kinney:2005ej,
	author = "Kinney, Justin and Maldacena, Juan Martin and Minwalla, Shiraz and Raju, Suvrat",
	title = "{An Index for 4 dimensional super conformal theories}",
	eprint = "hep-th/0510251",
	archivePrefix = "arXiv",
	doi = "10.1007/s00220-007-0258-7",
	journal = "Commun. Math. Phys.",
	volume = "275",
	pages = "209--254",
	year = "2007"
}

@article{Romelsberger:2005eg,
	author = "Romelsberger, Christian",
	title = "{Counting chiral primaries in N = 1, d=4 superconformal field theories}",
	eprint = "hep-th/0510060",
	archivePrefix = "arXiv",
	doi = "10.1016/j.nuclphysb.2006.03.037",
	journal = "Nucl. Phys. B",
	volume = "747",
	pages = "329--353",
	year = "2006"
}

@article{Dolan:2008qi,
	author = "Dolan, F. A. and Osborn, H.",
	title = "{Applications of the Superconformal Index for Protected Operators and q-Hypergeometric Identities to N=1 Dual Theories}",
	eprint = "0801.4947",
	archivePrefix = "arXiv",
	primaryClass = "hep-th",
	reportNumber = "DAMTP-08-07, DIAS-STP-08-02, SHEP-08-06",
	doi = "10.1016/j.nuclphysb.2009.01.028",
	journal = "Nucl. Phys. B",
	volume = "818",
	pages = "137--178",
	year = "2009"
}

@article{Pan:2025vyu,
	author = "Pan, Yiwen and Yang, Peihe",
	title = "{Exact non-Lagrangian Schur index in closed form}",
	eprint = "2509.20439",
	archivePrefix = "arXiv",
	primaryClass = "hep-th",
	month = "9",
	year = "2025"
}

@article{Gadde:2011ik,
	author = "Gadde, Abhijit and Rastelli, Leonardo and Razamat, Shlomo S. and Yan, Wenbin",
	title = "{The 4d Superconformal Index from q-deformed 2d Yang-Mills}",
	eprint = "1104.3850",
	archivePrefix = "arXiv",
	primaryClass = "hep-th",
	reportNumber = "YITP-SB-11-13",
	doi = "10.1103/PhysRevLett.106.241602",
	journal = "Phys. Rev. Lett.",
	volume = "106",
	pages = "241602",
	year = "2011"
}

@inbook{Rastelli:2014jja,
	author = "Rastelli, Leonardo and Razamat, Shlomo S.",
	editor = {Teschner, J{\"o}rg},
	title = "{The superconformal index of theories of class S.}",
	booktitle = "{New Dualities of Supersymmetric Gauge Theories}",
	eprint = "1412.7131",
	archivePrefix = "arXiv",
	primaryClass = "hep-th",
	doi = "10.1007/978-3-319-18769-3_9",
	pages = "261--305",
	year = "2016"
}

@article{Beem:2013sza,
	author = "Beem, Christopher and Lemos, Madalena and Liendo, Pedro and Peelaers, Wolfger and Rastelli, Leonardo and van Rees, Balt C.",
	title = "{Infinite Chiral Symmetry in Four Dimensions}",
	eprint = "1312.5344",
	archivePrefix = "arXiv",
	primaryClass = "hep-th",
	reportNumber = "YITP-SB-13-45, CERN-PH-TH-2013-311, HU-EP-13-78",
	doi = "10.1007/s00220-014-2272-x",
	journal = "Commun. Math. Phys.",
	volume = "336",
	number = "3",
	pages = "1359--1433",
	year = "2015"
}

@article{Zheng:2022zkm,
	author = "Zheng, Haocong and Pan, Yiwen and Wang, Yufan",
	title = "{Surface defects, flavored modular differential equations, and modularity}",
	eprint = "2207.10463",
	archivePrefix = "arXiv",
	primaryClass = "hep-th",
	doi = "10.1103/PhysRevD.106.105020",
	journal = "Phys. Rev. D",
	volume = "106",
	number = "10",
	pages = "105020",
	year = "2022"
}

@article{Pan:2023jjw,
	author = "Pan, Yiwen and Wang, Yufan",
	title = "{Flavored modular differential equations}",
	eprint = "2306.10569",
	archivePrefix = "arXiv",
	primaryClass = "hep-th",
	doi = "10.1103/PhysRevD.108.085027",
	journal = "Phys. Rev. D",
	volume = "108",
	number = "8",
	pages = "085027",
	year = "2023"
}

@article{Maruyoshi:2016tqk,
	author = "Maruyoshi, Kazunobu and Song, Jaewon",
	title = "{Enhancement of Supersymmetry via Renormalization Group Flow and the Superconformal Index}",
	eprint = "1606.05632",
	archivePrefix = "arXiv",
	primaryClass = "hep-th",
	reportNumber = "IMPERIAL-TP-16-KM-02",
	doi = "10.1103/PhysRevLett.118.151602",
	journal = "Phys. Rev. Lett.",
	volume = "118",
	number = "15",
	pages = "151602",
	year = "2017"
}

@article{Maruyoshi:2016aim,
	author = "Maruyoshi, Kazunobu and Song, Jaewon",
	title = "{$ \mathcal{N}=1 $ deformations and RG flows of $ \mathcal{N}=2 $ SCFTs}",
	eprint = "1607.04281",
	archivePrefix = "arXiv",
	primaryClass = "hep-th",
	reportNumber = "IMPERIAL-TP-16-KM-03",
	doi = "10.1007/JHEP02(2017)075",
	journal = "JHEP",
	volume = "02",
	pages = "075",
	year = "2017"
}

@article{Agarwal:2016pjo,
	author = "Agarwal, Prarit and Maruyoshi, Kazunobu and Song, Jaewon",
	title = "{$ \mathcal{N} $ =1 Deformations and RG flows of $ \mathcal{N} $ =2 SCFTs, part II: non-principal deformations}",
	eprint = "1610.05311",
	archivePrefix = "arXiv",
	primaryClass = "hep-th",
	reportNumber = "SNUTP16-006",
	doi = "10.1007/JHEP12(2016)103",
	journal = "JHEP",
	volume = "12",
	pages = "103",
	year = "2016",
	note = "[Addendum: JHEP 04, 113 (2017)]"
}

@article{Creutzig:2024ljv,
	author = "Creutzig, Thomas and Garner, Niklas and Kim, Heeyeon",
	title = "{Mirror symmetry and level-rank duality for 3d $\mathcal {N} = 4$ rank 0 SCFTs}",
	eprint = "2406.00138",
	archivePrefix = "arXiv",
	primaryClass = "hep-th",
	doi = "10.1007/s11005-025-02015-x",
	journal = "Lett. Math. Phys.",
	volume = "115",
	number = "6",
	pages = "123",
	year = "2025"
}

@article{Kontsevich:2008fj,
	author = "Kontsevich, Maxim and Soibelman, Yan",
	title = "{Stability structures, motivic Donaldson-Thomas invariants and cluster transformations}",
	eprint = "0811.2435",
	archivePrefix = "arXiv",
	primaryClass = "math.AG",
	month = "11",
	year = "2008"
}

@inproceedings{Mukhi:2019xjy,
	author = "Mukhi, Sunil",
	title = "{Classification of RCFT from Holomorphic Modular Bootstrap: A Status Report}",
	booktitle = "{Pollica Summer Workshop 2019}: {Mathematical and Geometric Tools for Conformal Field Theories}",
	eprint = "1910.02973",
	archivePrefix = "arXiv",
	primaryClass = "hep-th",
	month = "10",
	year = "2019"
}

@article{Mukhi:2019cpu,
	author = "Mukhi, Sunil and Poddar, Rahul and Singh, Palash",
	title = "{Contour integrals and the modular $ \mathcal{S} $-matrix}",
	eprint = "1912.04298",
	archivePrefix = "arXiv",
	primaryClass = "hep-th",
	doi = "10.1007/JHEP07(2020)045",
	journal = "JHEP",
	volume = "07",
	pages = "045",
	year = "2020"
}

@article{Mathur:1988rx,
	author = "Mathur, Samir D. and Mukhi, Sunil and Sen, Ashoke",
	title = "{Differential Equations for Correlators and Characters in Arbitrary Rational Conformal Field Theories}",
	reportNumber = "TIFR/TH/88-32",
	doi = "10.1016/0550-3213(89)90022-9",
	journal = "Nucl. Phys. B",
	volume = "312",
	pages = "15--57",
	year = "1989"
}

@article{Seiberg:1994rs,
	author = "Seiberg, N. and Witten, Edward",
	title = "{Electric - magnetic duality, monopole condensation, and confinement in N=2 supersymmetric Yang-Mills theory}",
	eprint = "hep-th/9407087",
	archivePrefix = "arXiv",
	reportNumber = "RU-94-52, IASSNS-HEP-94-43",
	doi = "10.1016/0550-3213(94)90124-4",
	journal = "Nucl. Phys. B",
	volume = "426",
	pages = "19--52",
	year = "1994",
	note = "[Erratum: Nucl.Phys.B 430, 485--486 (1994)]"
}

@article{Seiberg:1994aj,
	author = "Seiberg, N. and Witten, Edward",
	title = "{Monopoles, duality and chiral symmetry breaking in N=2 supersymmetric QCD}",
	eprint = "hep-th/9408099",
	archivePrefix = "arXiv",
	reportNumber = "RU-94-60, IASSNS-HEP-94-55",
	doi = "10.1016/0550-3213(94)90214-3",
	journal = "Nucl. Phys. B",
	volume = "431",
	pages = "484--550",
	year = "1994"
}

@article{Franc_Mason_2014, title={Fourier Coefficients of Vector-valued Modular Forms of Dimension 2}, volume={57}, DOI={10.4153/CMB-2014-007-3}, number={3}, journal={Canadian Mathematical Bulletin}, author={Franc, Cameron and Mason, Geoffrey}, year={2014}, pages={485–494}}

@article{Beem:2021zvt,
	author = "Beem, Christopher and Razamat, Shlomo S. and Singh, Palash",
	title = "{Schur indices of class S and quasimodular forms}",
	eprint = "2112.10715",
	archivePrefix = "arXiv",
	primaryClass = "hep-th",
	doi = "10.1103/PhysRevD.105.085009",
	journal = "Phys. Rev. D",
	volume = "105",
	number = "8",
	pages = "085009",
	year = "2022"
}

@article{Mukhi:2020gnj,
	author = "Mukhi, Sunil and Poddar, Rahul and Singh, Palash",
	title = "{Rational CFT with three characters: the quasi-character approach}",
	eprint = "2002.01949",
	archivePrefix = "arXiv",
	primaryClass = "hep-th",
	doi = "10.1007/JHEP05(2020)003",
	journal = "JHEP",
	volume = "05",
	pages = "003",
	year = "2020"
}

@article{Chandra:2025qpv,
	author = "Chandra, A. Ramesh and Mukhi, Sunil and Singh, Palash",
	title = "{Generalised 4d Partition Functions and Modular Differential Equations}",
	eprint = "2512.02107",
	archivePrefix = "arXiv",
	primaryClass = "hep-th",
	month = "12",
	year = "2025"
}

@article{Buican:2015hsa,
	author = "Buican, Matthew and Nishinaka, Takahiro",
	title = "{Argyres{\textendash}Douglas theories, S$^1$ reductions, and topological symmetries}",
	eprint = "1505.06205",
	archivePrefix = "arXiv",
	primaryClass = "hep-th",
	reportNumber = "RU-NHETC-2015-02",
	doi = "10.1088/1751-8113/49/4/045401",
	journal = "J. Phys. A",
	volume = "49",
	number = "4",
	pages = "045401",
	year = "2016"
}

@article{Buican:2015ina,
	author = "Buican, Matthew and Nishinaka, Takahiro",
	title = "{On the superconformal index of Argyres{\textendash}Douglas theories}",
	eprint = "1505.05884",
	archivePrefix = "arXiv",
	primaryClass = "hep-th",
	reportNumber = "RU-NHETC-2015-01",
	doi = "10.1088/1751-8113/49/1/015401",
	journal = "J. Phys. A",
	volume = "49",
	number = "1",
	pages = "015401",
	year = "2016"
}

@article{Buican:2017uka,
	author = "Buican, Matthew and Nishinaka, Takahiro",
	title = "{On Irregular Singularity Wave Functions and Superconformal Indices}",
	eprint = "1705.07173",
	archivePrefix = "arXiv",
	primaryClass = "hep-th",
	reportNumber = "QMUL-PH-17-XX",
	doi = "10.1007/JHEP09(2017)066",
	journal = "JHEP",
	volume = "09",
	pages = "066",
	year = "2017"
}
\end{document}